\documentclass[draft]{icrc}
\usepackage{fix2col}
\def\simge{\mathrel{    \rlap{\raise 0.511ex \hbox{$>$}}{\lower
0.511ex \hbox{$\sim$}}}}
\def\simle{\mathrel{    \rlap{\raise 0.511ex \hbox{$<$}}{\lower
0.511ex \hbox{$\sim$}}}}
\usepackage{times}
\usepackage{graphicx} 
\usepackage[bookmarks=false]{hyperref}
 \hypersetup{
   pdftitle={H.E.S.S. Lecture}
   pdfauthor={H. J. V\"olk}
   pdfkeywords{27th ICRC-Conference, Hamburg, 2001 }
   bookmarksopen={false}
  }
\begin{document}

\title{ Gamma-Ray Astronomy of Cosmic Rays
\\({\it Hess Lecture 2001})
\\
{\normalsize \mdseries For high resolution figures and a high quality manuscript refer to
\href{http://www.mpi-hd.mpg.de/hfm/HESS/HESS.html}{H.E.S.S webpage:}\\[0mm]
(http://www.mpi-hd.mpg.de/hfm/HESS/HESS.html),\\[-3mm]
and under H.E.S.S.-related
talks and publications, Conference proceedings, klick on Gamma-Ray
Astronomy of Cosmic Rays.}
}
\author[1]{Heinrich J. V\"olk}
\affil[1]{Max Planck Institut f\"ur Kernphysik,
Postfach 103980, D-69029 Heidelberg, Germany}

\correspondence{H.J. V\"olk \\(Heinrich.Voelk@mpi-hd.mpg.de)}

\runninghead{V\"olk: Gamma-Ray Astronomy of Cosmic Rays}
\firstpage{1}
\pubyear{2001}

\maketitle

\begin{abstract} Many of the basic problems in the astrophysics of charged Cosmic Rays remain on
principle unresolved by in situ observations in the Solar System due to the chaotic nature of the
propagation of these particles in Interstellar space. This concerns the existence and the nature
of localized individual particle sources as well as the transport in the Galaxy and establishes
the need for astronomical observations of secondary gamma-rays. The only exception may be the
highest energy particles at energies around $10^{20}$~eV which possibly reach us on straight line
orbits from their production sites. Recently such gamma-ray observations, both in space and on the
ground, have made great progress even though the instrumental sensitivities are still low. It is
argued that two basic questions, regarding first of all the Supernova Remnant source
hypothesis and secondly the contributions to the diffuse gamma-ray background, have come close to
an empirical resolution. Apart from motivations deriving from extragalactic astronomy this
expectation is at the root of the construction of a new generation of high-sensitivity gamma-ray
instruments. As a representative example the H.E.S.S. array of atmospheric Cherenkov telescopes is
described. \end{abstract}

\section{Introduction}

\noindent The International Cosmic Ray Conferences have made it a tradition
that a Lecture is given in honor of Victor Franz (Francis) Hess. This gives me
the opportunity to recount his discovery in 1912 which initiated the era of
Cosmic Ray research and paved the way for the development of particle physics.
The realization of the truly historic significance of this discovery appears to
be generally growing in the physics community. As part of this awareness it is
interesting to note recent efforts at the Universities of Innsbruck and Vienna
to emphasize and consolidate the memory of the activities of Hess in
Austria\footnote{See http://physik.uibk.ac.at/hephy/Hess/homepage/. This Web
site also contains an article in "Current Biography Yearbook 1963" which gives
an interesting and moving description of his life-story (Moritz, 1963).}.

I will then summarize our present knowledge about Ga\-lactic Cosmic Rays
(CRs) from in situ observations on the ground and in satellite experiments
above the atmosphere. In doing so I emphasize the lower energy range, below
the so-called "knee" of the energy spectrum near $10^{15}$~eV, because it contains the
overwhelming part of the CR energy density that determines the degree of
collective CR interaction with the rest of the Interstellar Medium. The
conclusion is, as we know, that the CRs in our local environment constitute
a relativistic nonthermal component of Interstellar Matter whose energy
density compares to all other relevant energy densities like the turbulent
and thermal energy density of the thermal gas and the Interstellar magnetic
field energy density. If true elsewhere in the Galaxy and beyond - and
there is every reason to assume this - then the CR component is an
significant dynamical element in Interstellar as well as in Intergalactic
space on sufficiently large spatial scales, and the question of the CR
sources becomes a matter of global astronomical importance.

\begin{figure}[t]
\figbox*{}{}{\includegraphics*[width=8.2cm]{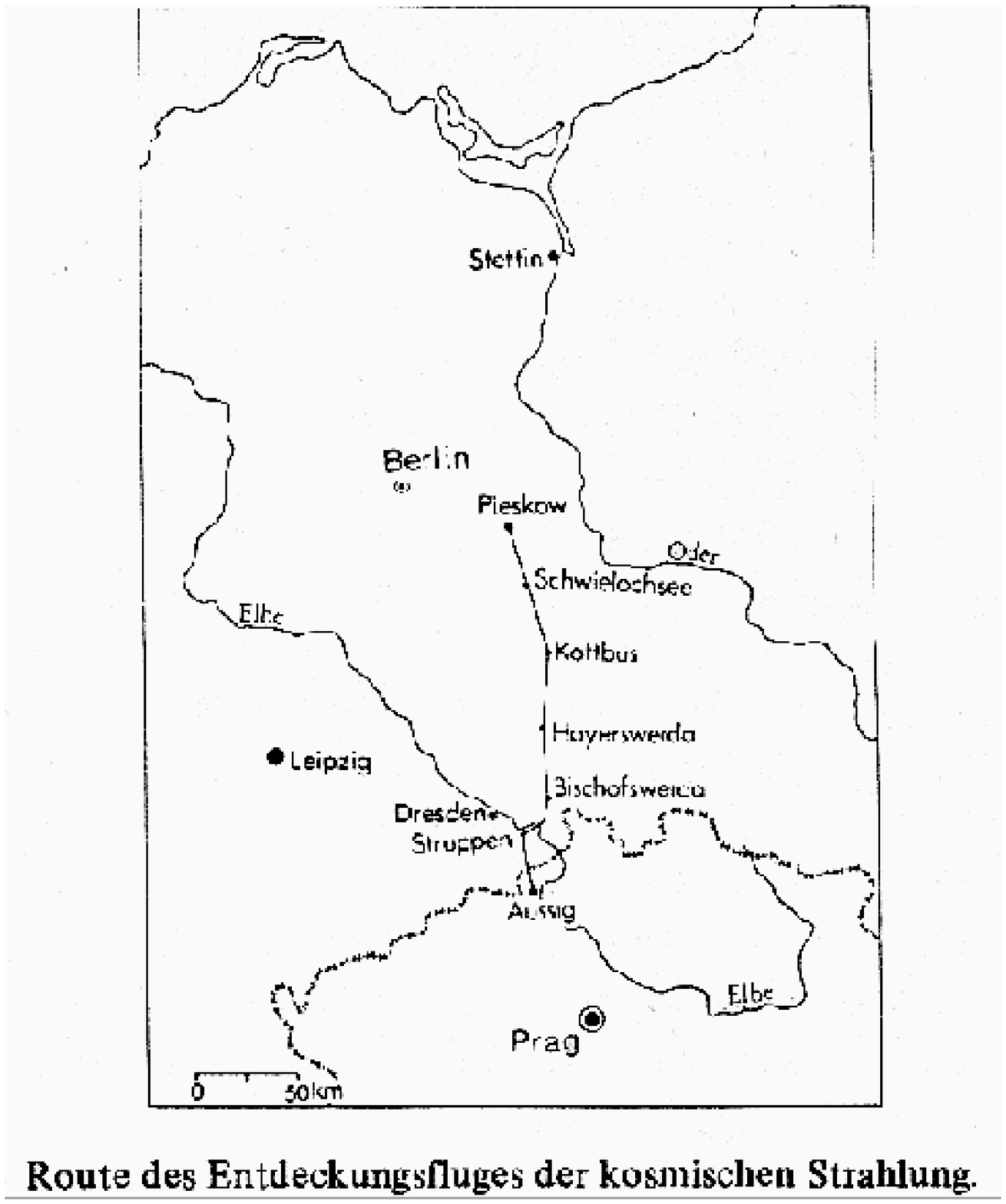}} 
\caption{The
flight on August 7, 1912, started in Aussig in then Bohemia at 06:12 and
reached the maximum height of 5350 m at 10:45. The landing took place near
Pieskow in Brandenburg at 12:15.}
\end{figure}

For these reasons we cannot discuss CR transport independently of the
dynamics of the interstellar gas and magnetic field. Basically this
nonlinear picture goes back to 1966 when Parker (1966) for the first time
connected the formation of large scale gas clouds in the Galactic disk with
the buoyant rise of the CRs in rarefied magnetized bubbles, proposing that
this should lead to their eventual escape to Intergalactic space.
Certainly, the process has to compete with diffusive escape in a CR-driven
Galactic Wind as emphasized more recently again by Breitschwerdt et al.
(1993) and Ptuskin et al. (1997) and it is difficult to estimate
quantitatively. But its importance for the development of CR astrophysics
is undisputed.


Attempts to verify the validity and the limitations of such a picture
require CR observations in the depth of the Cosmos, at least extending
across the Galaxy. This type of observations can in a direct form only be
done on secondary neutral particles, gamma-rays and neutrinos, whose
trajectories point back to their origin, and I shall discuss gamma-ray
observations here. 


My main topics will be recent gamma-ray investigations of presumed 
localized CR sources and of CR propagation in the Galaxy. This concerns first
of all the Supernova Remnant source hypothesis and the present
observational results in comparison with theoretical acceleration models. With
regard to CR transport and to the key question how representative the CR properties
near the Solar System are for the Galaxy as a whole, I will discuss the
unexpectedly hard spectrum of the diffuse gamma-ray flux from the Galactic disk,
its implications and possible explanation. The observed very small diffuse,
radial gamma-ray gradient in the Galactic disk poses another problem for CR
propagation theory. It may again be a nonlinear effect in the sense that an
increase of CR production due to a spatial concentration of sources leads to a
compensatory local velocity enhancement of the Galactic Wind which prevents the
expected increase of the local particle density.

\begin{sloppypar}
  Accompanied and partly driven by theoretical develop\-ments gamma-ray observations of Galactic and
Extragalactic sources have made significant progress over the last years, despite the rather low
detector sensitivities. This has encouraged the construction of a new generation of larger and
more sensitive instruments. I will conclude this lecture with a description of a representative
experiment, the ground-based stereoscopic H.E.S.S. array of imaging atmospheric Cherenkov
telescopes. It is named in honor of Victor Hess and is being built in Namibia.
\end{sloppypar}

\section{The time of Victor Hess}

\noindent In his balloon flights, Hess brought sealed ionization
chambers into the upper atmosphere to measure the rate of ionization induced
as a function of height. From about 1500~m above ground the ionization
increased monotonically with height. In his highest flight on August 7, 1912
(Fig. 1,2), at the maximum altitude of 5300 m the ionization rate increased by
a factor of two relative to the ground which could not have come from
radioactive material on the ground or in the air.

\begin{figure}[htbp]
\figbox*{}{}{\includegraphics*[width=8.2cm]{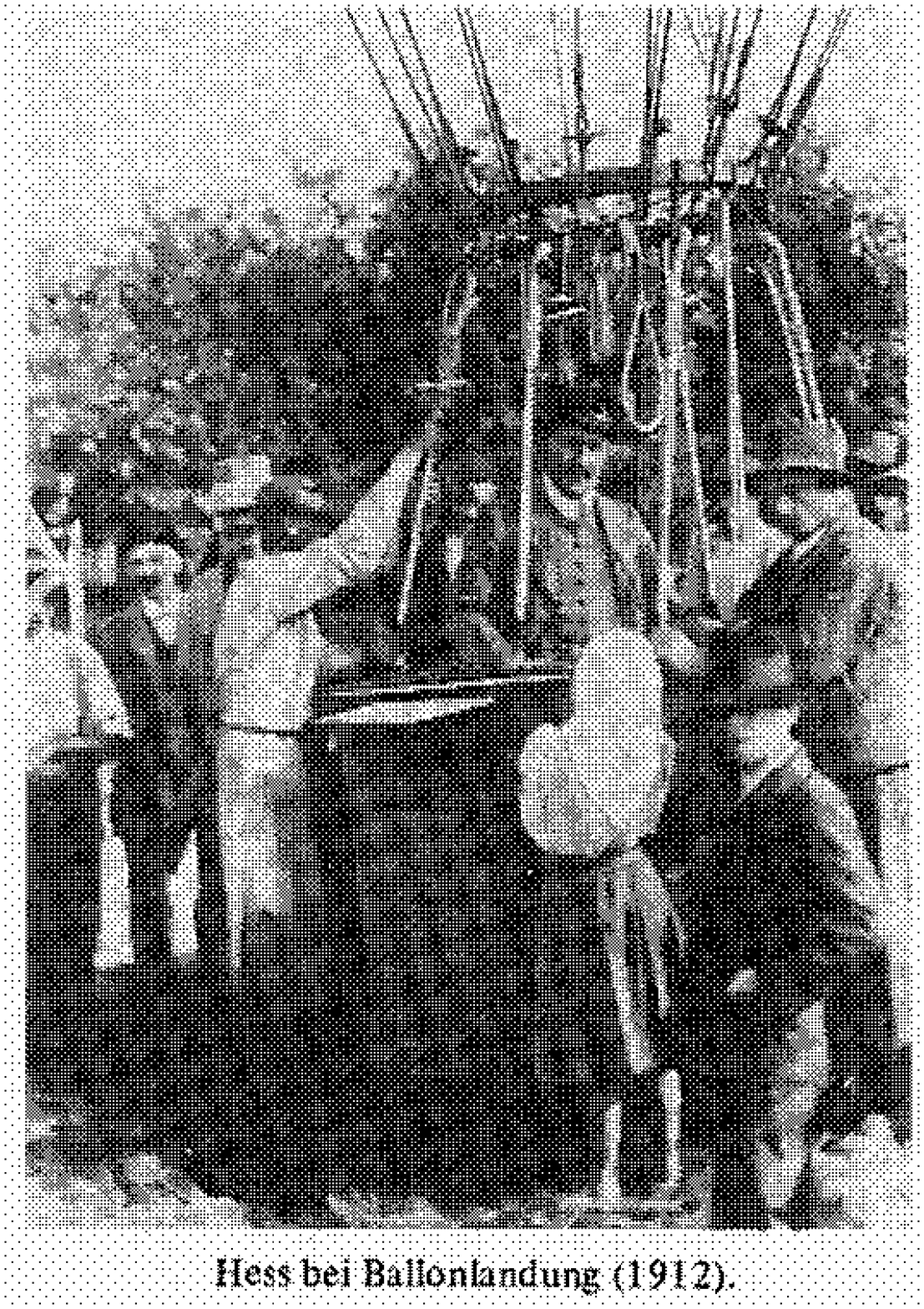}} 
\caption{Victor
Hess (in the middle) and his crew in the balloon gondola after the landing
in Pieskow.} 
\end{figure}

\begin{sloppypar}
The flight took about 6 hours and after the subsequent checks the balloon
crew could return to Vienna by night train. In 1912 the geographical world
of experiments was still small!
\end{sloppypar}

\begin{figure*}[t]
\figbox*{}{}{\includegraphics*[width=17.7cm,height=13cm]{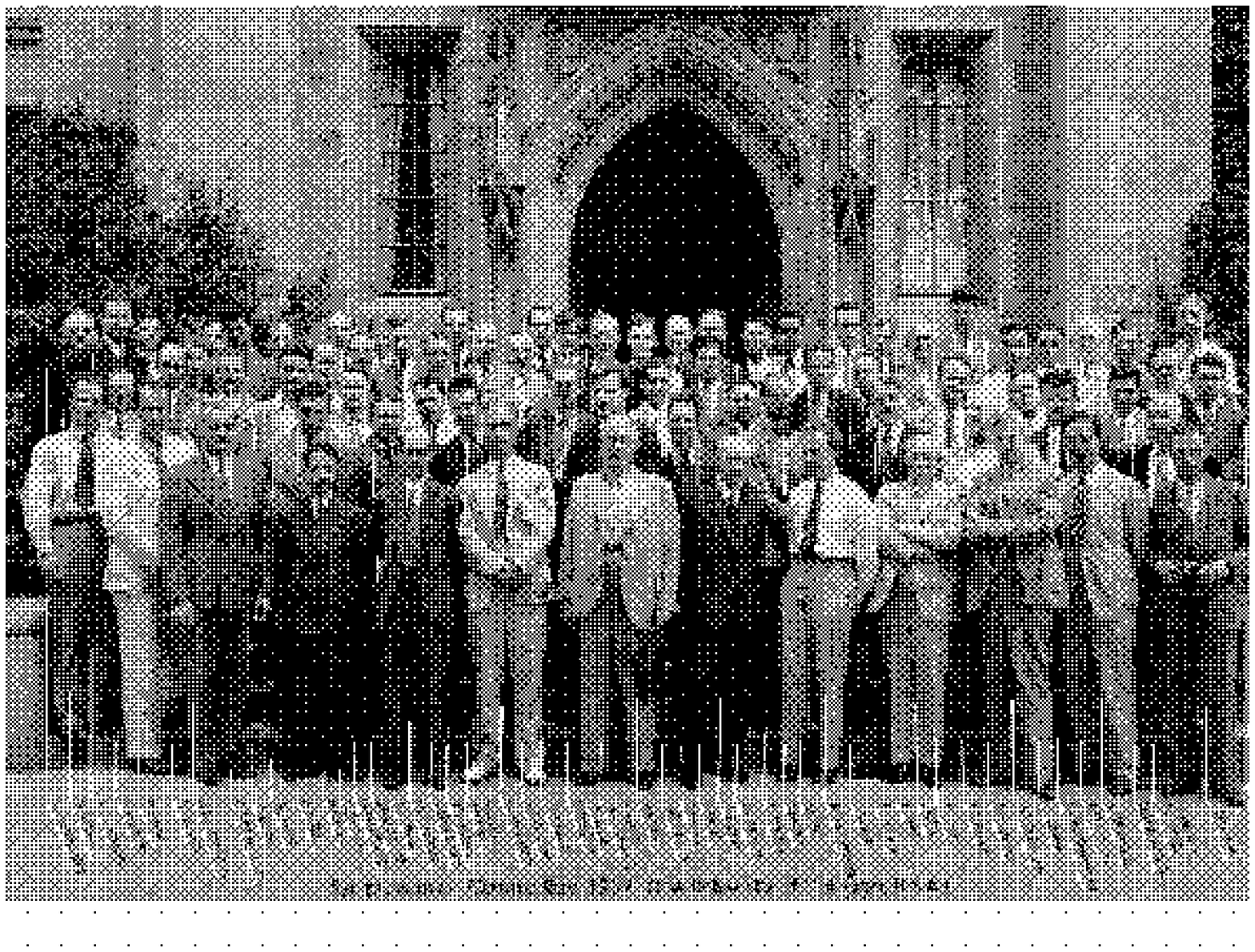}}  
\caption{The participants of the Symposium on Cosmic Rays, Chicago 1939.
A number of individuals are identified by name on the bottom of the
figure by the author (archive Max-Planck-Institut f\"ur Krenphysik).} 
\end{figure*}

The conclusion that Hess (1912) could finally draw was clear and
surprising: "The results ... appear most likely explainable by the
assumption that a radiation of very high penetration power enters our
atmosphere from above\footnote{In the German original: "Die Ergebnisse
... scheinen am ehesten durch die Annahme erkl\"art werden zu k\"onnen,
dass eine Strahlung von sehr hoher Durchdringungskraft von oben her in
unsere Atmosph\"are eindringt."}". Subsequent balloon measurements at even
much greater altitudes, in particular by W. Kohlh\"orster, confirmed
and strengthened the result.
 
The history of the later, difficult and sometimes controversial
investigations into the nature of these very energetic particles is a
fairly long one. Eventually, after twenty years of experimentation by many
groups and the development of important new techniques, it was found that
the Cosmic Rays above the atmosphere are mostly positively charged nuclear
particles, in fact mainly protons. These particles became vital tools for
the emerging field of particle physics and all of the new particles
discovered until the beginnings of the accelerator era in the mid-fifties
were found in CR interactions. Fig.~3 is a photograph of the participants
in the {\it Symposium on Cosmic Rays} held 1939 in Chicago, showing Victor
Hess in the center together with many other well-known physicists: Werner
Heisenberg, Walter Bothe, Arthur Holly Compton, Robert Oppenheimer, Edward
Teller, Pierre Auger, Carl Anderson, Hans Bethe, and others. I got this
picture originally from Professor Maurice Shapiro who is at the conference
here in Hamburg: you can see him in the upper left hand corner as a young
graduate student.

\section{Cosmic Rays near the Solar System} 
\subsection{Primary particle energy spectra, energetics}

\noindent Direct measurements of CRs are made nowadays with sophisticated
detector combinations on satellites, balloons, and (at energies above
$10^{15}$ eV/nucleon) on the ground. The resulting energy spectra are
similar for many chemical elements considered to be primary, i.e. directly
accelerated with little subsequent nuclear transformations. There is also a
primary CR electron component and at GeV energies its flux is two orders of
magnitude below that for protons. Most impressively, the all-particle
energy spectrum extends over more than 11 orders of magnitude in energy. It
is the prototype of a nonthermal spectrum, without a sign for a
characteristic energy scale (Fig. 4).

The differential energy spectrum is approximately a power law in energy $E$
$\propto E^{-2.75}$ beyond the range of influence from the Sun, for
$10^{10}~{\rm eV} <E< {\rm few}~10^{15}$~eV. The
spectral features at several $10^{15}$~ and $10^{18}$~eV,
respectively, may indicate different particle sources, or alternatively,
different energy dependences of the propagation conditions in the separate
energy regions. The corresponding estimate of the CR energy density $E_{CR}$
in the neighborhood of the Solar System is of the order of
$1$~eV/cm$^{-3}$, about equal to the thermal energy density $E_{gas}$ of
the Interstellar gas as well as the typical Interstellar magnetic energy
$E_{mag}$, measured by other means: $E_{CR} \sim E_{gas} \sim E_{mag}$. I
have discussed the significance of this equality before.

\begin{figure}[htbp]
\figbox*{}{}{\includegraphics*[width=8.2cm]{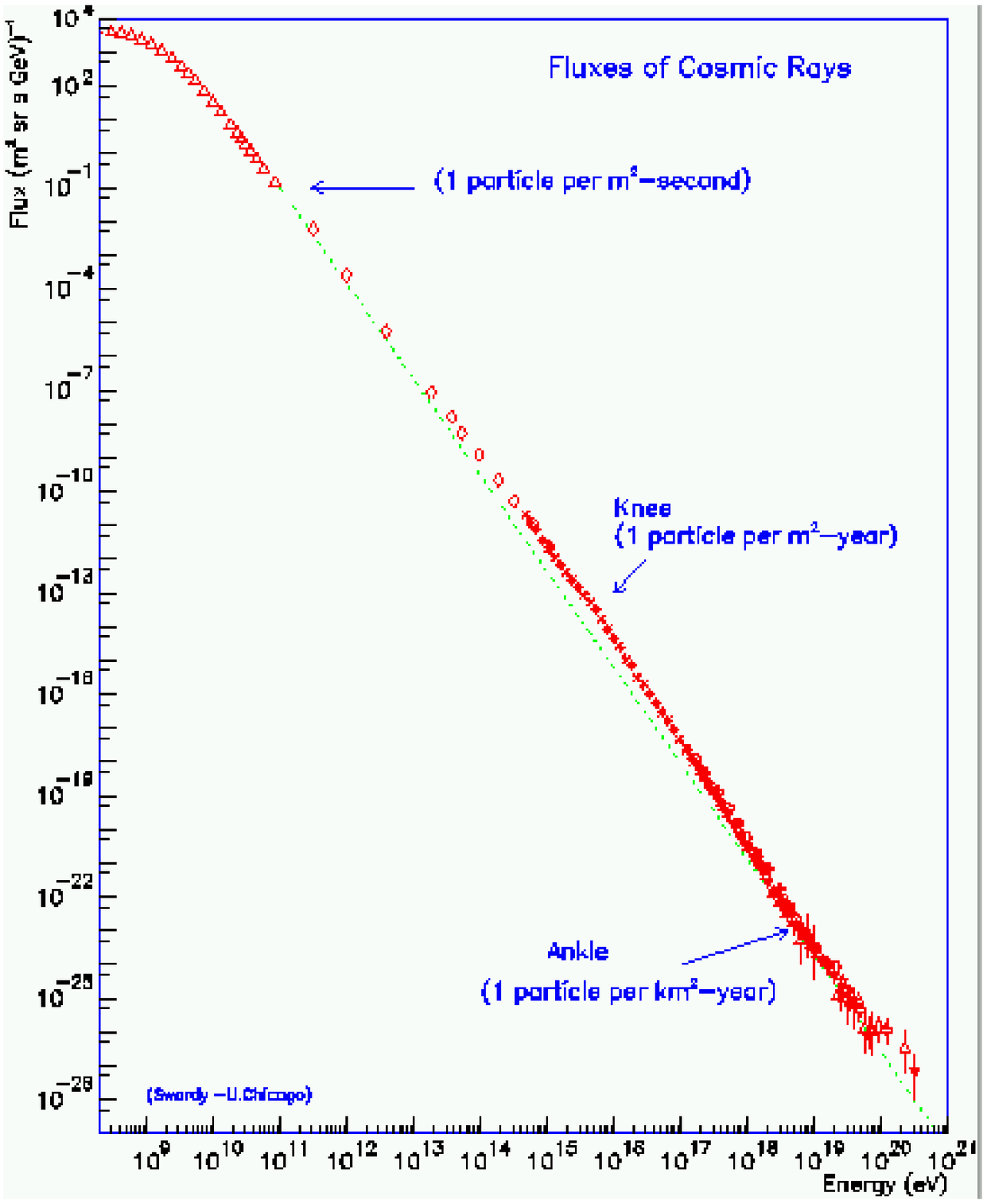}} 
\caption{The
all-particle CR energy spectrum. For $E\simge 10$~GeV, the spectrum is a
power law, slightly steepening at a ${\rm few}~10^{15}$~eV, the
so-called Knee, and hardening at a ${\rm few}~10^{18}$~eV, 
the so-called Ankle. (Adapted
from Cronin et al. 1997; courtesy S. Swordy.)} 
\end{figure}

How should we picture the overall energy distribution of the thermal gas plus the CRs as they
coexist in a given volume element in space? This question has no unique answer because, even if
the gas and the CRs are energized at the same place in a cosmic accelerator like a Solar Flare or
a Supernova Remnant, their spatial propagation can be very different. Nevertheless, at such an
accelerator, the nonthermal power law distribution of the CRs should grow out of the thermal
distribution somewhere above the gas thermal energy. This can be seen in Fig. 5 which shows an
analytical calculation of diffusive particle acceleration at a shock wave. The example also
indicates the relative energetics: despite the fact that the particle number density of the gas
exceeds that of the CRs by three orders of magnitude, the inverse is roughly true for the mean
particle energies. Therefore such a process can indeed lead to approximately equal energy
densities of the two components.

\begin{figure}[htbp]
\figbox*{}{}{\includegraphics*[width=8.2cm,height=7.8cm]{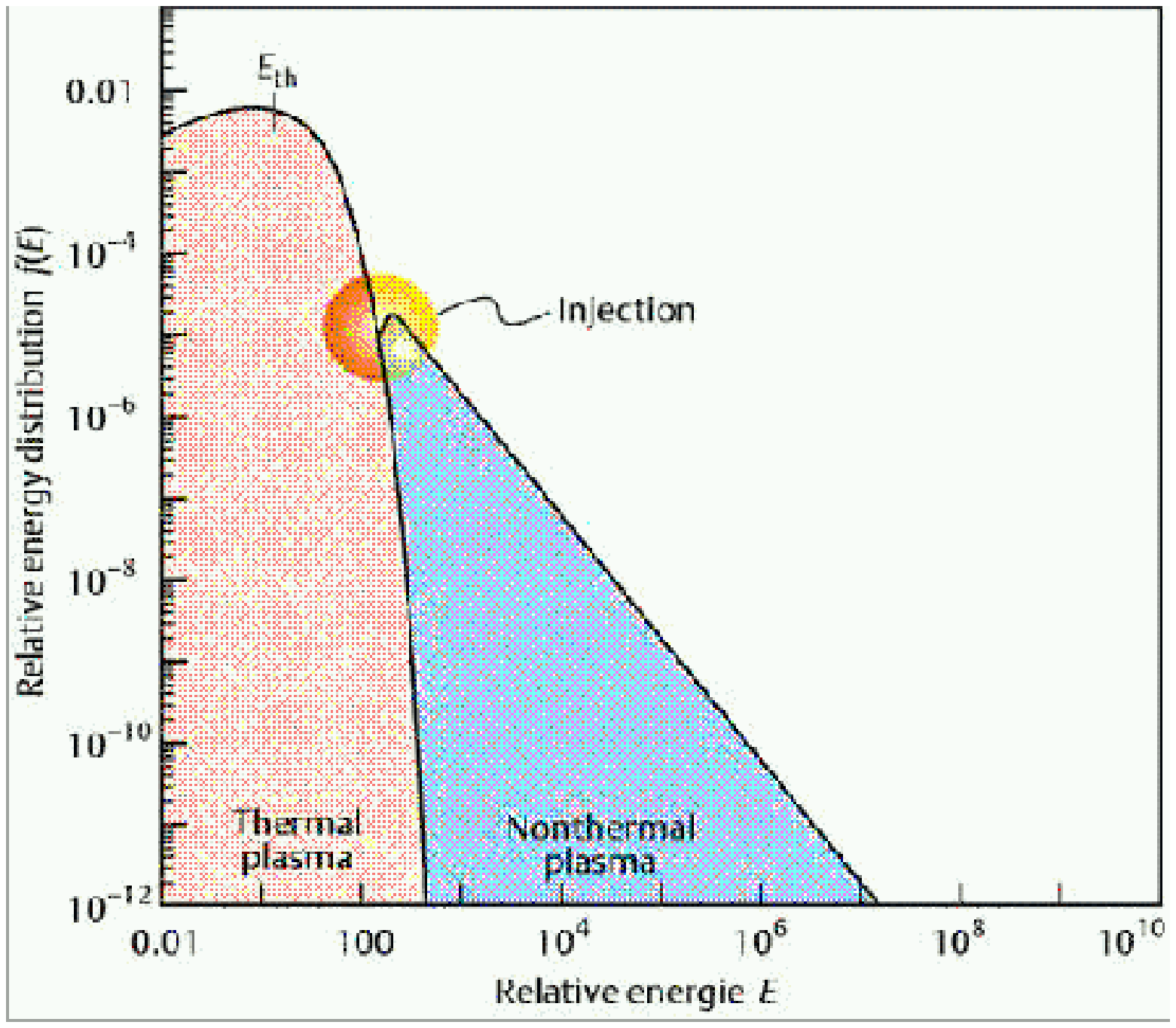}}
\caption{Total
energy distribution of thermal plasma (gas) plus nonthermal plasma (CRs)
near a diffusively accelerating shock wave: the thermal (Maxwellian) energy
distribution joins rather smoothly to the nonthermal power law CR
distribution at an "injection" energy that is several times larger than the
mean thermal energy $E_{th}$. Only supra-thermal particles above this
injection energy can participate in the collective acceleration process
(adapted from Malkov \& V\"olk, 1998; courtesy "Sterne und Weltraum").}
\end{figure}

\subsection{Cosmic Ray source spectra, composition}

\noindent The observed CR energy spectra are not necessarily identical with
the spectra of the particles as they are emitted from their sources. The
connection between the two is rather given by the particle propagation
properties. Observations show that the ratio between the energy spectra of
CR spallation products and their primary particles decreases with
energy (Fig.~6). For energies above $10$~GeV/nucleon this translates
directly into a corresponding energy dependence of the average amount
of Interstellar matter "seen" by CR particles. If we assume the particles to
be produced deeply inside the dense Galactic gas disk then this implies a
shorter residence time there for higher energy particles than for those of lower energy
before they eventually escape to Intergalactic Space.

\begin{figure}[htbp]
\figbox*{}{}{\includegraphics*[width=8.2cm]{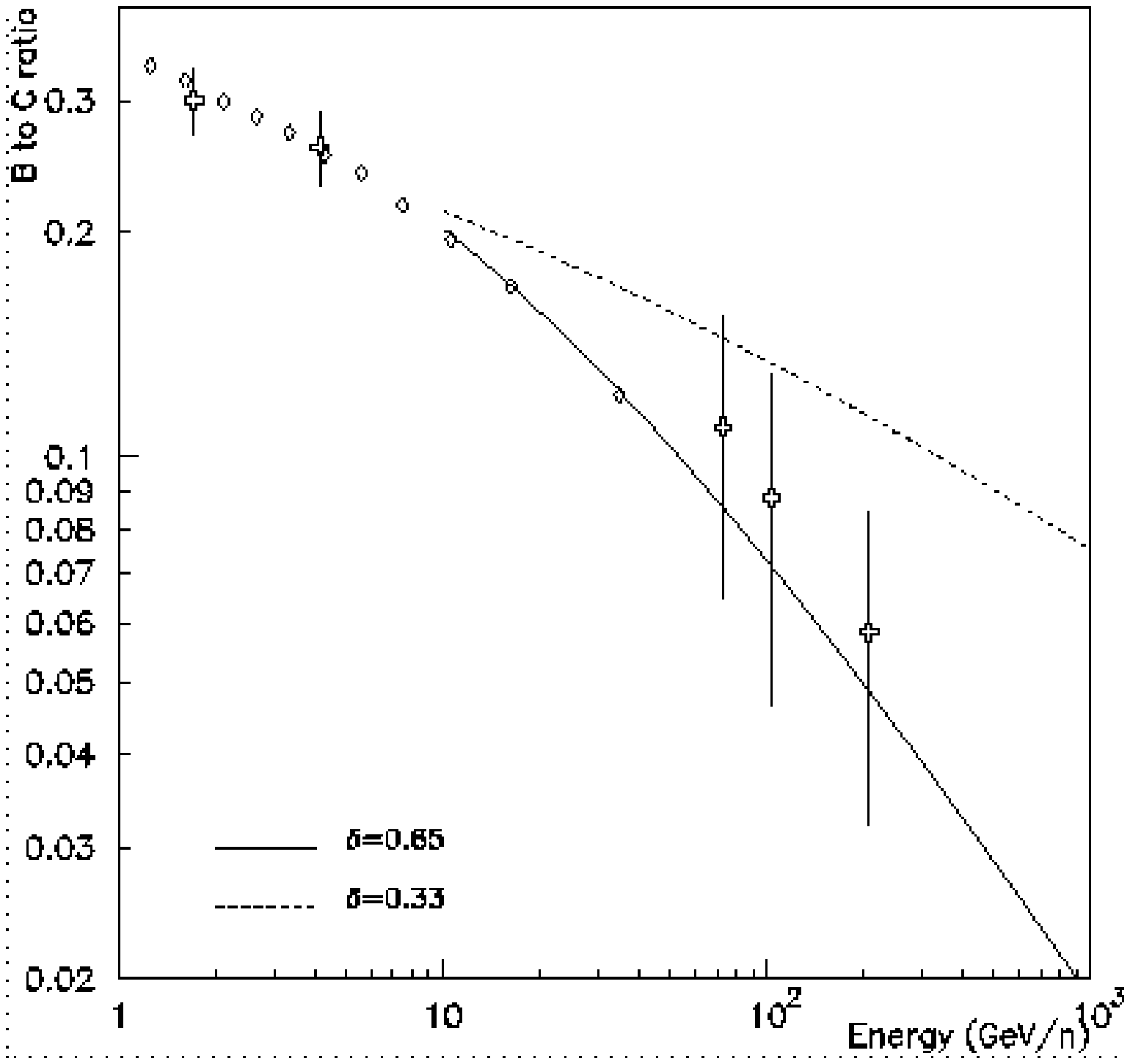}} 
\caption{The measured ratio of (secondary) Boron to (primary) Carbon
nuclei as a function of energy/nucleon up to $200$~GeV/nucleon:
diamond-shaped points from Engelmann et al. (1990), crosses from Swordy et
al. (1990). Two different power law dependences $\propto (E/{\rm
nucleon})^{- \delta}$ are indicated for comparison. (Courtesy S. Swordy.)}
\end{figure}

Let us now in addition take the particle sources as well as the particles
released from them to be uniformly distributed across the Galactic
disk that includes also the Solar System where we measure. Then one can
infer from such local measurements that the source spectra for various
different elements are very hard: $dN/dE \propto (E/{\rm nucleon})^{-2.1}$
at least up to $E/{\rm nucleon} \sim 10^3$~GeV/nucleon, i.e. up to the TeV
region ($10^3~{\rm GeV} = 1$~TeV).

\begin{sloppypar}
The chemical composition at the sources is much less known than the
energy spectra. However, up to an energy of $100$~GeV/nucleon the source material
corresponds to rather normal Interstellar Medium material (gas and dust),
with a number of characteristic deviations (see below). At higher energies,
no determination of source spectra has been possible until now. From all we
know, in the arriving CRs heavier nuclei appear to become more
abundant with increasing energy, also beyond the knee. Above the ankle
virtually nothing is known about the composition; the arriving particles
could even be gamma-rays. It is clear that this is a wide open field and
several new experiments with sophisticated instrumentation are operating or
are being built at present. They will be among the main topics at this
conference.
\end{sloppypar}

\section{The need for gamma-ray observations}

\noindent The most fundamental questions in CR astrophysics concern the individual sources - if
such localized accelerators exist in the first place - and the transport of these
ultrarelativistic particles in the Galaxy. Due to the deviation of charged particle orbits from a
linear path in the chaotic interstellar magnetic fields, localized accelerators cannot be
identified by observation of such particles near the Earth, as indicated in the cartoon Fig.~7. In
the same sense also the propagation in the Galaxy remains unknown, apart from the energy
dependence of the average matter density which the CRs have seen. Therefore there is a need for
astronomical observations that give directional and, with the aid of multi-wavelength
observations, also distance information.

\begin{figure}[htbp]
\figbox*{}{}{\includegraphics*[width=8.2cm,height=5.4cm]{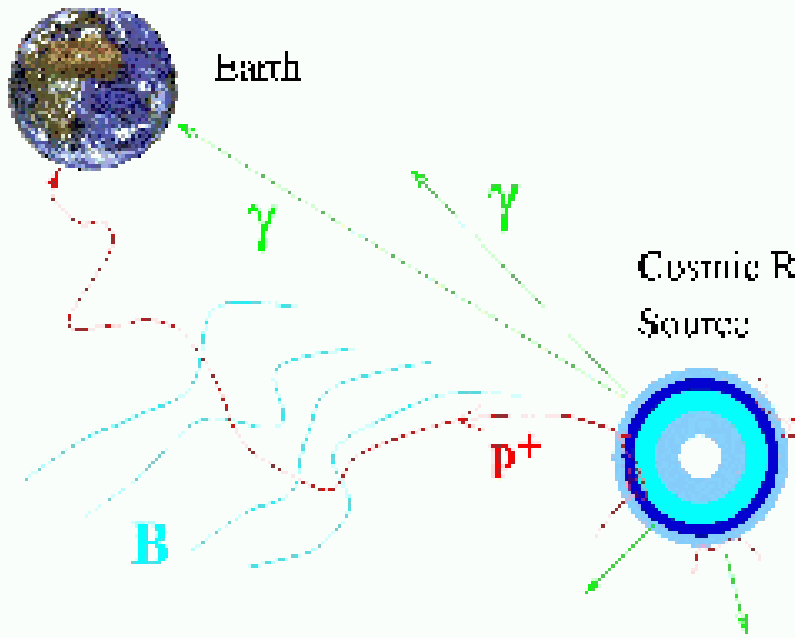}}
\caption{Charged particles $p^+$ emitted from a localized Cosmic Ray source
have followed chaotic orbits in the interstellar magnetic field $B$ when they
reach the Earth. Therefore their arrival direction does not point back to the
source as do $\gamma$-rays. (Courtesy G. P\"uhlhofer.)} 
\end{figure}

In a direct form this can be done with high energy ($\simge 0.1$~GeV) $\gamma$-rays or with 
neutrinos
as neutral interaction products from CR collisions with thermal gas atoms (pion-decay, electron
Bremsstrahlung), or Inverse Compton effect on photon fields, mainly the microwave background
radiation. Integrated along the line of sight, the Inverse Compton $\gamma$-ray emission from CR
electrons turns out to be comparable in magnitude with that from nucleon-induced pion decay, given
typical interstellar gas densities. Since such secondary $\gamma$-rays and neutrinos already have
energies in the range of interest for the sources of the charged CRs, they portray the high energy
CRs at first hand. Obviously the program is then the following: with the aid of $\gamma$-ray or
neutrino observations individual CR sources may be found from amongst the regions of localized
emission. Under this assumption of source localization the spatial and spectral dependence of the
diffuse $\gamma$-ray and neutrino emission is then determined by the transport properties of the 
CRs
outside the sources.

Apart from these direct astronomical signals there is a further indicator of
growing importance. This is high frequency synchrotron emission, sometimes
reaching hard X-ray energies. But the inference on the radiating energetic
electrons is less direct; secondary information like polarization or a
clearly nonthermal frequency spectrum must be available to distinguish this
signal from Bremsstrahlung emission by electrons of comparable energy. In
fact, most energetic objects like Pulsars, Supernova Remnants, or entire
galaxies, emit radio synchrotron radiation which is of great value to
ascertain the spatial extent of the CR distribution and of the magnetic
field, as well as their equipartition strengths. Beyond that, the
synchrotron interpretation of seemingly nonthermal hard X-ray emission from
several Pulsars and Supernova Remnants has been taken as an indication that
electrons in the tens of TeV energy range are present in these objects. To
the extent that this is the case, there is little doubt that also CR
nucleons should have been accelerated.

The field of radio or X-ray observations is not my topic here. But I shall
make use of their results in discussing the search for the CR sources.
Neutrino astronomy at high energies $\simge 1$~GeV is still in the R \& D
phase. Therefore I will concentrate here on $\gamma$-ray astronomy. Above
$0.1$~GeV the dominant instruments are directionally sensitive
$\gamma$-ray detectors on satellites above the atmosphere and, for energies
$\simge 10$~GeV, ground-based imaging atmospheric Cherenkov telescopes.
There exist also interesting non-imaging $\gamma$-ray detectors using Solar
power plants. They will presumably enhance the ground based capabilities in
the future.

\begin{figure}[t]
\figbox*{}{}{\includegraphics*[width=8.2cm]{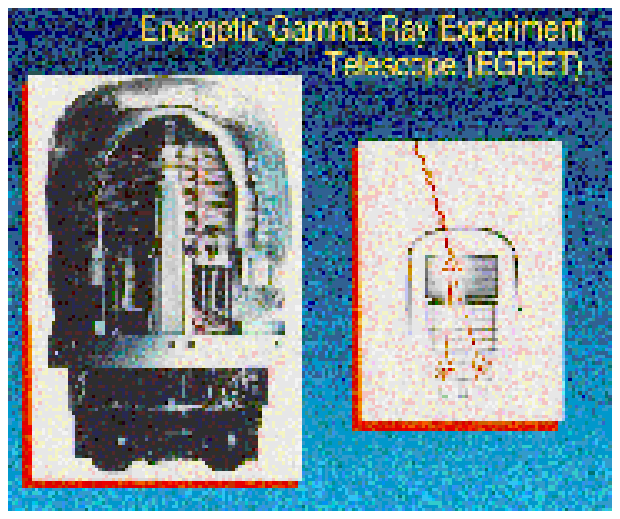}}
\caption{Cut-away of the Energetic Gamma Ray Experiment Telescope (EGRET) on
the Compton Gamma Ray Observatory. The track of the pair produced by the
interaction of the primary $\gamma$-ray in the upper part of the detector is
made visible in an adjacent spark chamber (right panel). The detector is
surrounded by an anti-coincidence hood where, in contrast to $\gamma$-rays,
charged CRs produce a scintillator signal (left panel).} 
\end{figure}

\section{Gamma-ray instruments}


\noindent At the energies concerned, pair production is the dominant process
for $\gamma$-ray absorption in matter. In satellite instruments the pair is
directly observed (Fig. 8).

In addition, a satellite detector can be furnished with an anti-coincidence
shield to effectively discriminate against the dominant background of
charged CRs. Previous instruments on the satellites SAS-II, Cos B, and CGRO
were characterized by a large field of view $\Delta \Omega$ of order $\pi$,
and a relatively small effective area below $1~{\rm m}^2$ for
$\gamma$-ray
energies between $30~{\rm MeV}$ and a few~GeV.

\begin{sloppypar}
Ground-based imaging atmospheric Cherenkov telescopes on the other hand
(Fig. 9) detect the Cherenkov light from the secondary electrons and
positrons of the electromagnetic shower in the atmosphere that is produced
by the primary cosmic $\gamma$-ray, with a maximum particle density at a
height of about $10$~km. The most important instruments are the Whipple,
CANGAROO (Collaboration of Australia and Nippon for a GAmma Ray Observatory
in the Outback), CAT (Cherenkov Array at Themis), and HEGRA (High Energy
Gamma Ray Astronomy) telescopes. The strong background due to the Cherenkov
emission from hadronic showers caused by charged CR nuclei is largely
removed by analysis of the image in the telescope's focal plane detector.
This is possible because hadronic showers are typically much broader than
electromagnetic showers, and not even approximately rotationally symmetric around the shower axis. 
For the background reduction
it is therefore extremely valuable to use a stereoscopic method in which
several telescopes are viewing the same shower in coincidence from
different directions within the $\sim 1$~degree Cherenkov light cone (Fig.
10). Such a land surveyor's technique also allows the determination of the
arrival direction and of the $\gamma$-ray energy on an event by event
basis. The instruments are generally characterized by a small field of view
of a few degrees, and a very large effective area of the order of
$10^5~{\rm m}^2$, with threshold energies of a few 100~GeV up to TeV
energies at present.
\end{sloppypar}

\begin{figure}[t]
\figbox*{}{}{\includegraphics*[width=8.2cm]{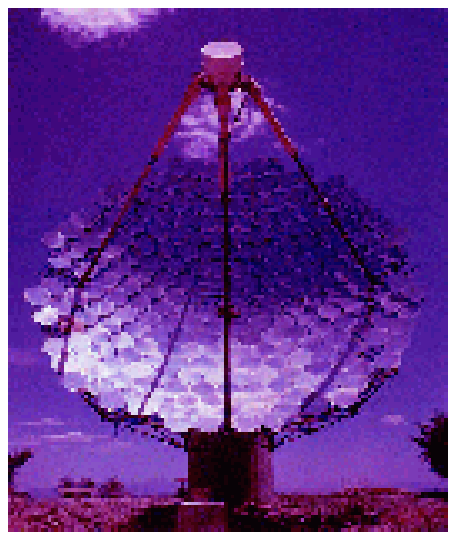}} 
\caption{Photograph
of the pioneering Whipple imaging atmospheric Cherenkov telescope in Arizona,
USA. With a diameter of $10$~m it has been until recently the largest operating
imaging telescope. The tessellated mirror consists of $248$ spherically shaped
hexagonally cut mirrors, each $61$~cm across. In the focus a fast camera whose
pixels are photomultipliers records the shower image against the night sky
background (Cawley et al. 1990).} 
\end{figure}

A well-known example of an almost entirely nonthermal source is the Crab
Nebula, powered by the relativistic wind of electron-positron pairs from
the Crab Pulsar (Fig. 11). In the $\gamma$-ray regime the satellite
instruments are assumed to document the transition from the synchrotron
radiation at lower energies to the Inverse Compton emission that dominates
for $E_{\gamma}$ above 1~GeV and is eventually detected by the Cherenkov
detectors.

\begin{figure}[t]
\figbox*{}{}{\includegraphics*[width=8.2cm]{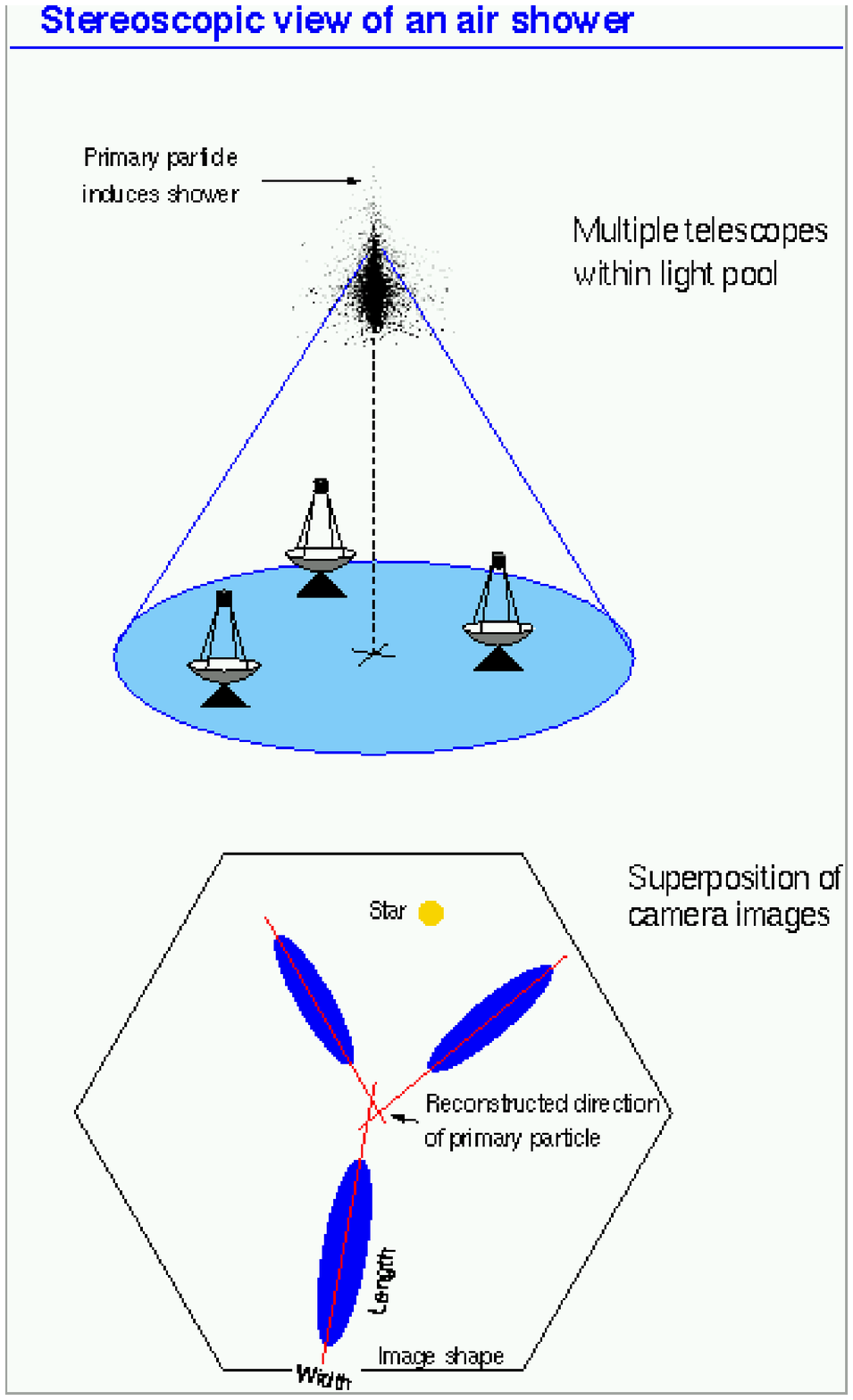}}
\caption{Stereoscopic $\gamma$-ray observation by several telescopes
positioned in the Cherenkov cone of an atmospheric shower. The generally
elliptical images, superposed into one of the focal plane cameras, determine
the arrival direction through the intersection point of their extrapolated
major axes. In analogous fashion also the impact point of the shower axis on
the ground is determined. This method was first introduced in the HEGRA
stereoscopic system.} 
\end{figure}

\section{The Supernova Remnant (SNR) source hypothesis}

\noindent Since the proposal by Baade and Zwicky (1934) Supernova
explosions are the primarily considered candidates for the CR sources up to
the knee of the spectrum\footnote{In their seminal paper Baade and Zwicky
also assumed for Supernova explosions "... that a considerable part of the
total radiation $E_T$ is emitted in the form of very hard rays or energetic
particles", and advanced the view "... that a super-nova represents the
transition of an ordinary star into a {\it neutron star}, consisting mainly
of neutrons.".}. The argument from today's point of view concerns diffuse
Supernova Remnants, not Pulsars nor Black Holes, and consists of three
elements:

1. As an ensemble, Supernova Remnants imply the largest
mechanical energy
release $(dE/dt)_{SN}$ into the Interstellar Medium which is available as
free energy for particle acceleration and heating of the thermal gas:
\begin{eqnarray*}
(dE/dt)_{SN} & = & \nu_{SN}\times E_{SN, mech} \sim 10^{42} \rm{erg/s} \\ 
             &   & \times \nu_{SN}/(1/30yr) \times E_{SN, mech}/(10^{51}\rm{erg}) \\
             &   & \sim 10 \times (dE/dt)_{CR},
\end{eqnarray*}
\begin{sloppypar}
\noindent  
in the Galaxy, where $\nu_{SN}$ and $E_{SN}$ denote the Galactic Supernova
rate and the mechanical energy output per SN, respectively, and
$(dE/dt)_{CR}$ is the total amount of energy leaving the Galaxy in the form
of CRs. Thus SNRs can indeed supply the very large amount of energy
required even though not by a large margin.
\end{sloppypar}

2. Diffusive shock acceleration at the leading outer SNR shock leads to a
hard momentum spectrum, approximately $\propto p^{-2}$. Given the above
requirement of high efficiency, acceleration must be a nonlinear dynamic
process which indeed it is.

3. The expected elemental composition of the accelerated component is basically
given by that of average interstellar matter - somewhat enriched by progenitor
wind material - swept up by the Supernova blast wave. A smaller contribution is
expected from ejected material of the progenitor stars if it is either
accelerated directly by secondary processes or has modified the circumstellar
gas of neighboring stars, exploding subsequently. This is essentially what the
measurements require.

\begin{figure*}[t]
\figbox*{}{}{\includegraphics*[width=16.cm,height=12cm]{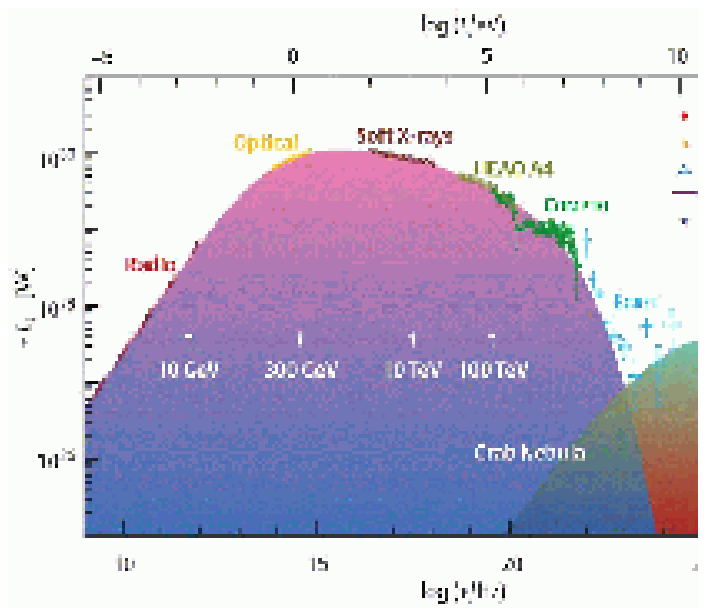}}
\caption{Spectral energy distribution of the emission from the Crab Nebula.
The electron energies producing the dominant synchrotron peak at lower
energies are indicated by the arrows. The CGRO $\gamma$-telescopes COMPTEL
and EGRET determine the synchrotron fall-off and the transition to the
Inverse Compton peak expected at some tens of GeV, and indicated by the
Cherenkov telescope measurements. (Adapted from Aharonian \& Atoyan (1998); 
Courtesy "Sterne und Weltraum".)}
\end{figure*}

There may be a theoretical problem with the SNR origin hypothesis, since the
cutoff energy predicted by quasilinear acceleration models lies typically an
order of magnitude below the "knee". However, I do not think that this is an
unsurmountable obstacle. The theory is still incomplete, despite the
sophistication it has reached in the last years. And there are reasons to
believe that a full consideration of the strongly nonlinear plasma physics of
the acceleration process will remove the deficiency.

All in all the arguments are quite persuasive. Yet this is still only
theory and we better turn to observations. They present a complex picture.

\subsection{Detected SNRs}

\noindent The EGRET experiment on the Compton Gamma Ray Observatory has
detected many Galactic $\gamma$-ray sources. Up to now the majority is still
not identified with known astronomical objects.

Also several SNRs with typical diameters smaller than a $1^{\circ}$ have
been detected, but the $\gamma$-rays are not firmly due to SNR shock
acceleration. They might as well be rather due to Pulsars or Pulsar
Nebulae. The identification of SNRs in the GeV energy region is in fact
very difficult, due to the steep spectrum of the strong diffuse
$\gamma$-ray background from the CRs in the Interstellar Medium. This steep
fall-off of the diffuse background spectrum combined with the assumed hard
source spectra suggests significantly better detection capabilities at
higher energies, in particular in the TeV range. As a consequence ground
based Cherenkov telescopes appear more suitable detectors, although the
present generation of instruments is only marginally sensitive for the
detection of SNRs. This is even true for luminous, very nearby objects.

Up to now three such TeV-detections have been reported (Fig. 12, 13,
14). They are on the $5\sigma$~level and still need confirmation by
independent groups:

(a) SN 1006 and SNR RX J1713.7-3946 in the Southern Hemisphere (CANGAROO
telescope)

(b) Cassiopeia~A in the Northern Sky (HEGRA telescope system).

\begin{figure}[htbp]
\figbox*{}{}{\includegraphics*[width=8.2cm]{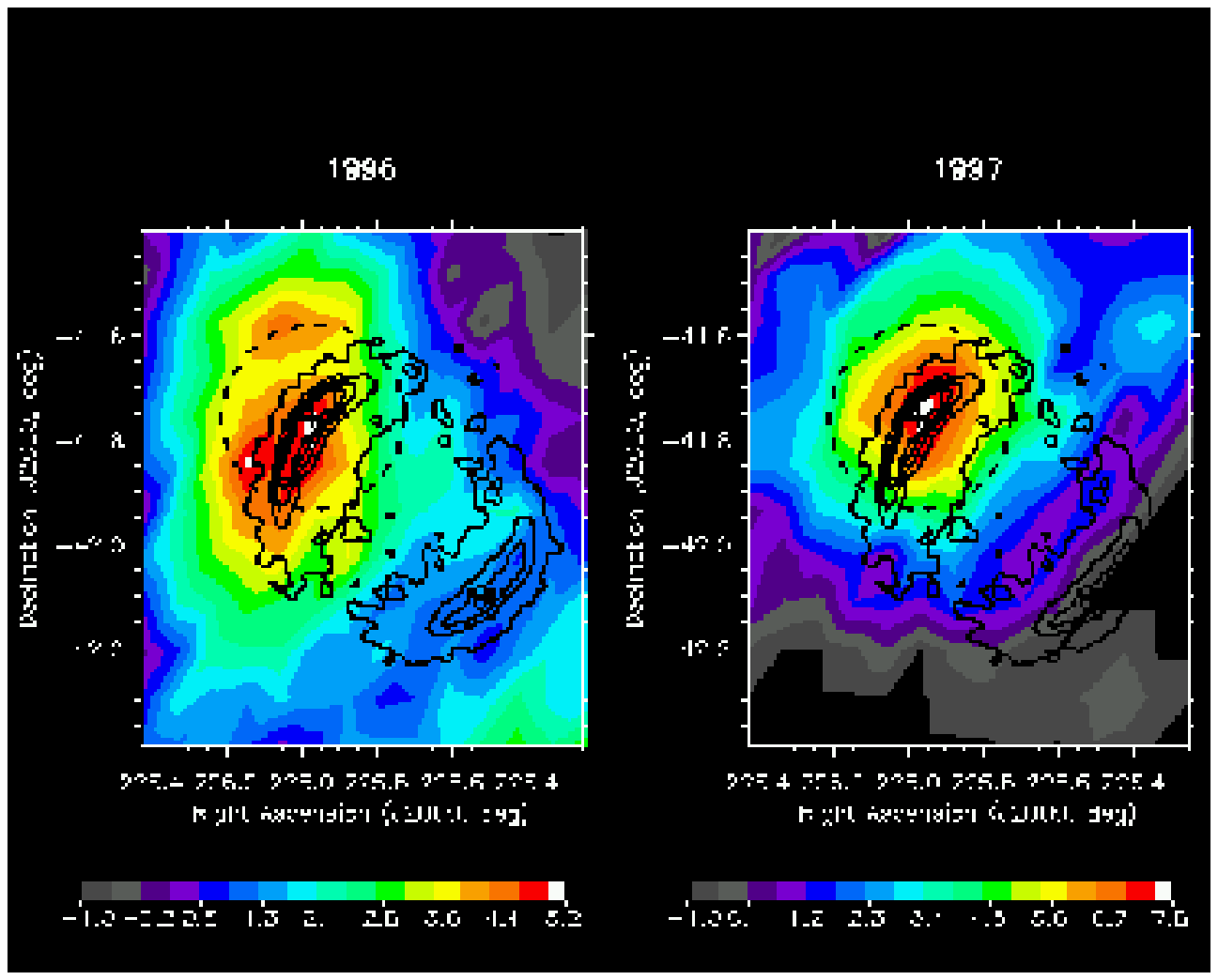}}
\caption{TeV $\gamma$-ray significance map from CANGAROO (Tanimori et al.
1998) for SN 1006, superposed on the nonthermal X-ray intensity contours
obtained by Koyama et al. (1995) with the ASCA satellite.}
\end{figure}

From published results, there exist basically single flux values around 1~TeV, with minimum
information on the spectrum. All three sources have also been detected in nonthermal X-rays,
supposedly synchrotron radiation from multi-TeV electrons. With simple assumptions the
$\gamma$-ray flux can be made consistent with Inverse Compton emission of such high energy
electrons. This is rather persuasive given the fact that the X-ray emission of both Southern
sources is even dominated by the nonthermal emission. In particular for SN 1006, a SN type Ia in a
uniform Interstellar Medium, with a possibly low gas density, a low magnetic field strength $B
\approx 6 \mu$Gauss allows consistency between the X-ray synchrotron emission on the one hand, and
Inverse Compton TeV $\gamma$-ray radiation on the Cosmic Microwave Background on the other - by
the same electrons - consistent with a differential spectral index of $-2$. A qualitatively
similar conclusion has been drawn for the emission from SNR RX J1713.7-3946, and more results are
expected at this conference from the new 10~m CANGAROO-II telescope. This does not claim that
there are no hadronic $\gamma$-rays, but it says that they may simply not be needed to explain the
reported $\gamma$-ray results.

\begin{figure}[htbp]
\figbox*{}{}{\includegraphics*[width=8.2cm]{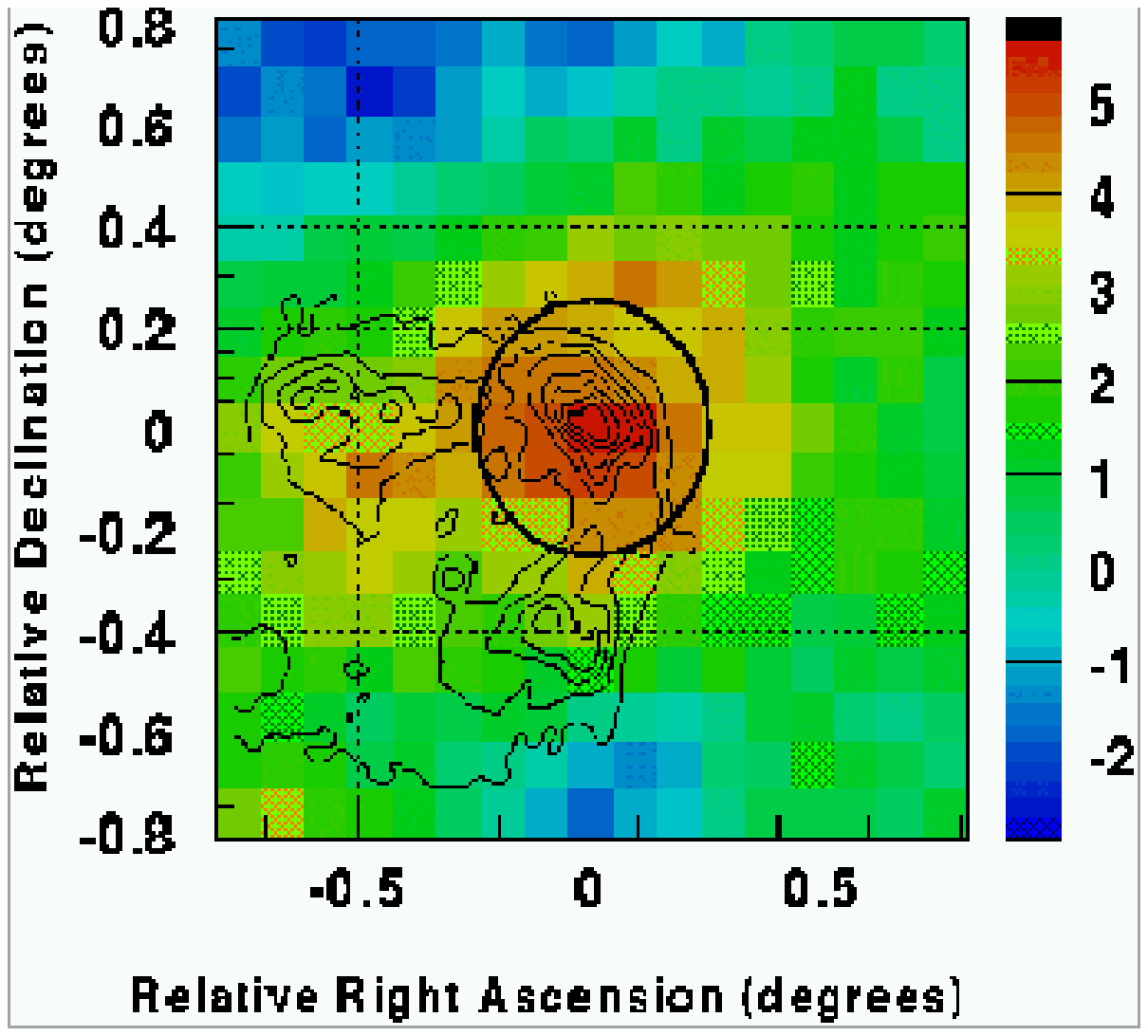}}
\caption{TeV $\gamma$-ray significance maps from CANGAROO (Muraishi et al.
2000), superposed on the nonthermal X-ray intensity contours for SNR RX
J1713.7-3946 from ASCA observations by Koyama et al. (1997).}
\end{figure}

This has led some researchers to a far reaching and pessimistic conclusion.
It says that the TeV $\gamma$-rays from such individual objects are
probably due to Inverse Compton collisions of CR electrons with background
photons alone and, in particular, that SNRs are not the sources of the CR
nuclei as well. I believe that such conclusions stem rather more from an
overly optimistic interpretation of the early models for the hadronic
emission expected from these objects (Drury et al. 1994; Naito \& Takahara
1994). As a consequence, now the disappointment is equally exaggerated.

\begin{sloppypar}
Cassiopeia~A is a somewhat different case (Fig.~14). In contrast to the two previous objects it is
a $\gamma$-ray point source for all practical considerations. Physically it is a quite nonuniform
source as judged from radio observations. In fact, Cassiopeia~A is believed to be the result of an
explosion into the nonuniform wind structure of a Wolf-Rayet star progenitor that went through
several very different mass loss phases. Detailed modeling of the observed radio and X-ray
synchrotron spectra, and of the resulting electron Bremsstrahlung and Inverse Compton emissions
from this high density object is again consistent with an electronic origin of the $\gamma$-ray
emission. On the other hand, one can make plausible estimates of the CR nucleon density in this
remnant and they suggest a strong hadronic $\gamma$-ray component of the same order. From Fig. 14
the estimated electronic spectrum falls off rather strongly with energy already at $1$~TeV,
whereas there is no obvious reason for the hadronic spectrum to follow this behavior. Yet the
observational spectral index estimate has a sufficiently large statistical error so that no
statistically significant exclusion of an electronic origin of the $\gamma$-ray emission is
possible from such arguments.

\begin{figure}[htbp]
\figbox*{}{}{\includegraphics*[width=8.2cm]{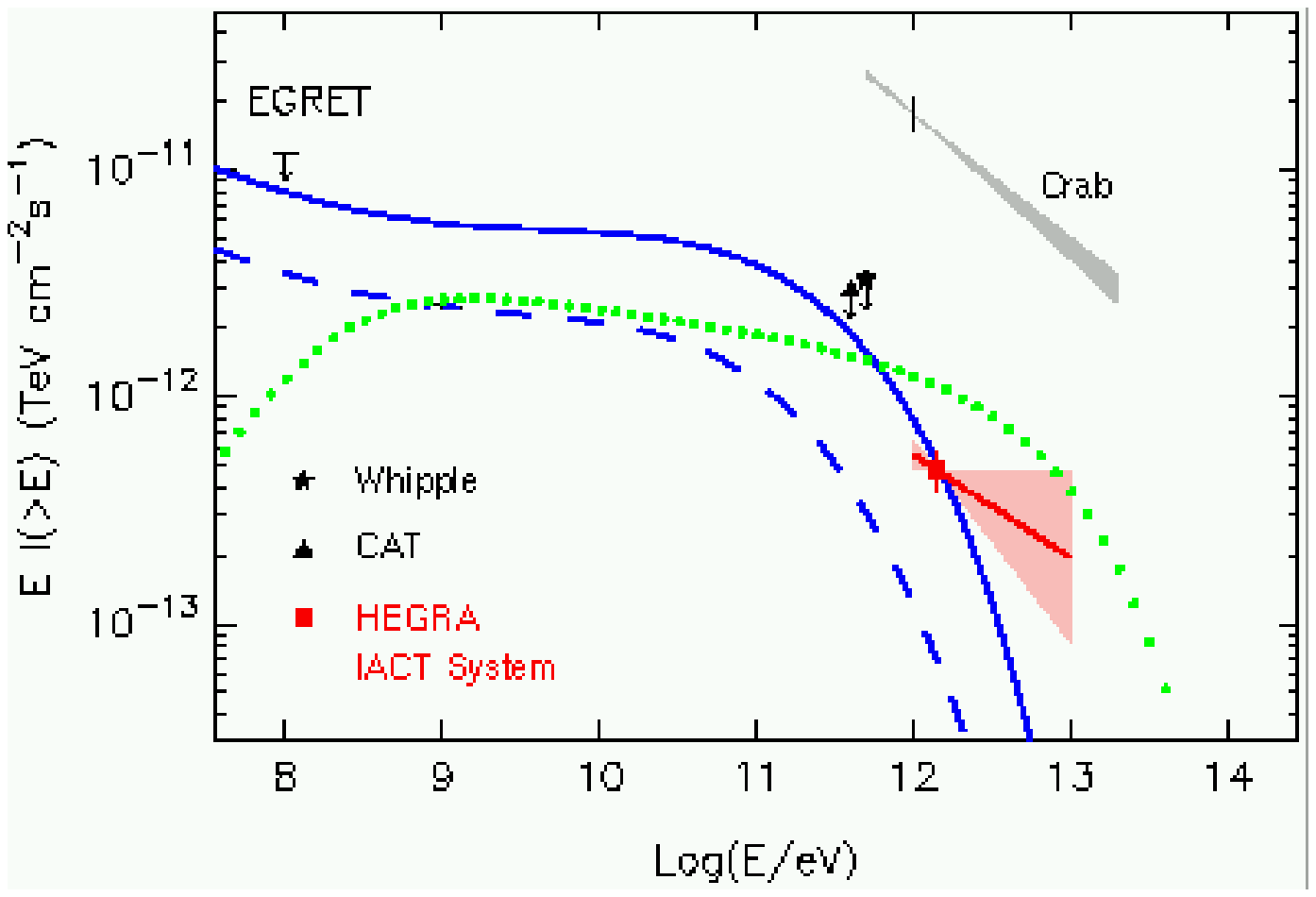}}
\caption{The HEGRA $\gamma$-ray flux from Cassiopeia~A (Aharonian et al.
2001a), together with the upper limits from EGRET on the one hand and the
Whipple and CAT Cherenkov telescopes on the other, in relation to several
model calculations. The solid and dashed curves show the electronic spectrum
for two different sets of parameters; the dotted curve is an estimate for
the hadronic spectrum. The observed spectral index of the HEGRA flux is
similar to that for the Crab Nebula spectrum, within the indicated large $1
\sigma$ uncertainty.} 
\end{figure}

A refinement of the theory in terms of time-dependent kinetic solutions,
simultaneously for electronic and hadronic emission, gives for SN 1006 
comparable hadronic and Inverse Compton emission. A distinction will require a
measurement of the $\gamma$-ray energy spectrum at high energies $0.1 \simle
E \simle 10$~TeV, i.e. above and below the expected Inverse Compton cutoff. For Cassiopeia~A, a 
similar spherically symmetric acceleration calculation results in a sizeable overproduction of 
hadronic $\gamma$-rays at 1 TeV. This could have several reasons, all connected with the poorly 
known circumstellar environment of the explosion site. Only spectroscopic measurements will help 
to unravel its dynamical structure.
\end{sloppypar}
 
\subsection{The case of Tycho's Supernova}

\noindent In the Northern Hemisphere Cassiopeia~A is the brightest Galactic shell-type SNR at
radio wavelengths. And it is the youngest one known, with an age of about 320 years. A seemingly
much simpler object in our close vicinity is the remnant of Tycho's Supernova from 1572 A.D. The
progenitor is thought to be an accreting White Dwarf - finally pushed over the Chandrasekhar
limit- in an otherwise rather uniform Interstellar Medium. It is also not dominated by nonthermal
electrons, showing instead a rich X-ray line spectrum. Thus it looks like an excellent candidate
for hadronic $\gamma$-ray emission, at least for energies above 1~TeV, never mind a possible
Inverse Compton component.

In a rather deep observation with the HEGRA telescope system, Tycho was not
detected (Fig. 15). However, the upper limit on the TeV flux is essentially near
the predicted hadronic flux from spherically symmetric kinetic theory, if this
prediction is renormalized to the expected reduction of the injection efficiency
for a remnant in a uniform external magnetic field. A ten times more sensitive
instrument should readily detect it, unless the theory is wrong.

\begin{figure}[htbp]
\figbox*{}{}{\includegraphics*[width=8.2cm]{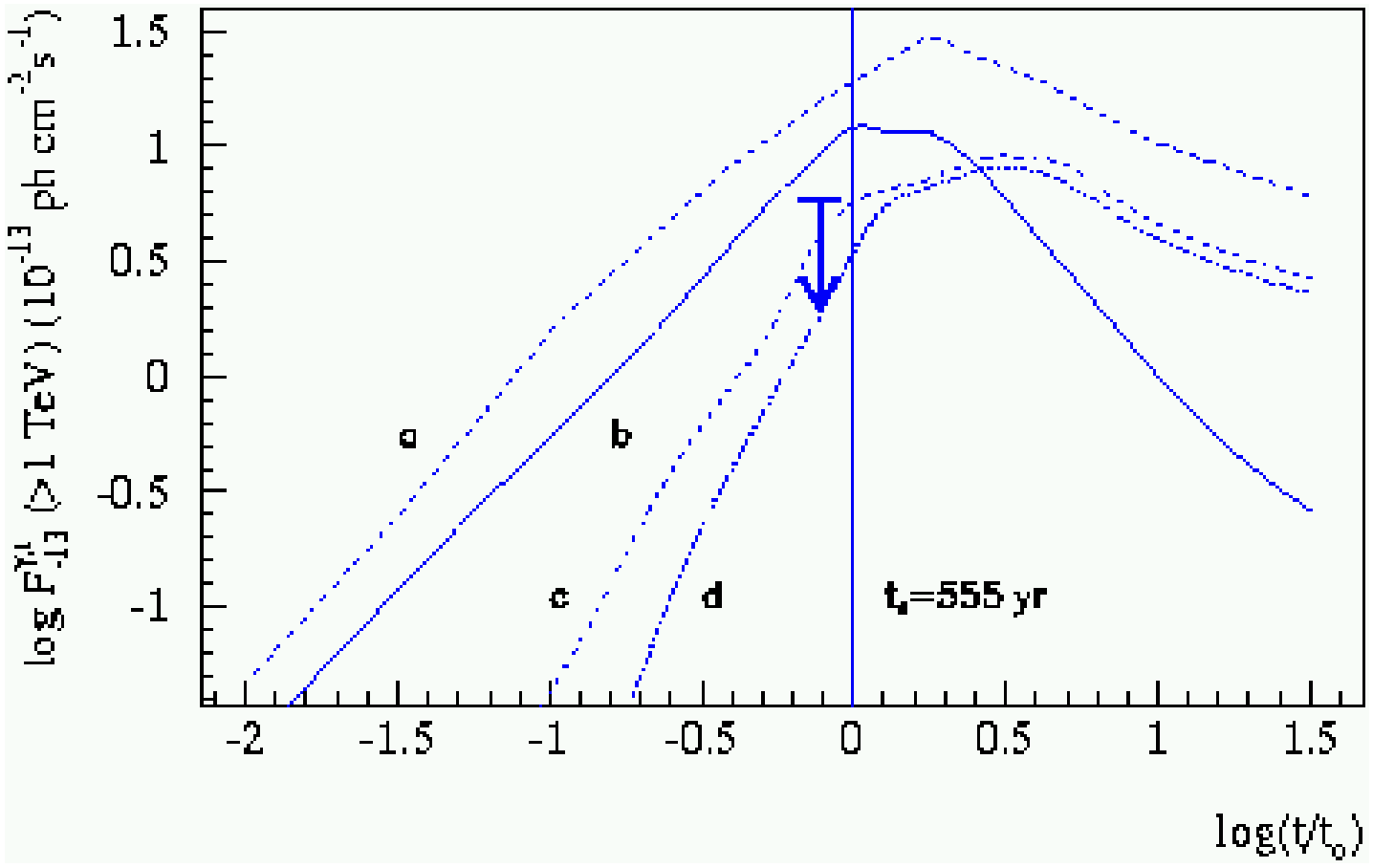}}
\caption{The predicted TeV $\gamma$-ray flux from Tycho's SNR as a function
of age, renormalized by a factor $1/3$, for various assumptions (a), (b), (c)
about the nucleon injection rate into the acceleration process at the outer
SNR shock, and stellar ejecta velocity distribution (d), together with the
HEGRA upper limit (Aharonian et al. 2001b). The favored case (c),
corresponding to the injection of $1$~ in $10^4$~ of the incoming upstream
nuclei, lies closely above the observational upper limit.} 
\end{figure}

\subsection{Conclusions}

\noindent What conclusions should we draw from this discussion? I would like
to summarize them in four points:

1. There are less than a handful of direct shell SNR source detections up to now and they need to
be confirmed. In particular the spectral information is still insufficient.

2. Theoretical models of diffusive shock acceleration are consistent with a roughly equal mixture
of $\pi^0$-decay and Inverse Compton fluxes at energies around 1~TeV. Present $\gamma$-ray
observations are not in contradiction with theory. The object Cassiopeia~A must be considered a 
special case due to its complex circumstellar dynamics.

3. At present there exists a tantalizing observational uncertainty about the hadronic $\gamma$-ray
fluxes. However, this uncertainty should be removed by the generation of experiments presently
under construction, one way or the other. The capabilities on the ground (at high energies $\simge
50$~GeV) and in space (at lower energies, up to some tens of GeV) will allow morphological studies
of nearby extended SNRs. In combination they will give $\gamma$-ray spectral coverage up to $\sim
10$~TeV. Inverse Compton emission will either be harder than any conceivable hadronic emission, or
it will have a comparable spectral slope but a low cutoff. In addition the number of detected
sources should increase by an order of magnitude, giving statistical weight to these distinctions.

4. Nevertheless, the result might not prove the SNR source hypothesis. And then we would have to
ask ourselves the difficult question, where we should turn to. Presumably only the much less
well-understood Galactic jet sources would be left: X-ray Binaries, Microquasars, or even Galactic
gamma-ray bursters. It is hard to imagine a more interesting and challenging situation for CR
physics!

\section{Diffuse Galactic gamma-ray emission (Transport)}

\noindent Let me now come to CR propagation in the Galaxy, i.e. the
transport properties of the nonthermal Interstellar component and to discuss two unexpected 
features of the Galactic emission that have been discovered some years ago. They are probably the 
main anomalies, given the limited spectral coverage and angular resolution of present instruments.

\begin{figure}[htbp]
\figbox*{}{}{\includegraphics*[width=8.2cm]{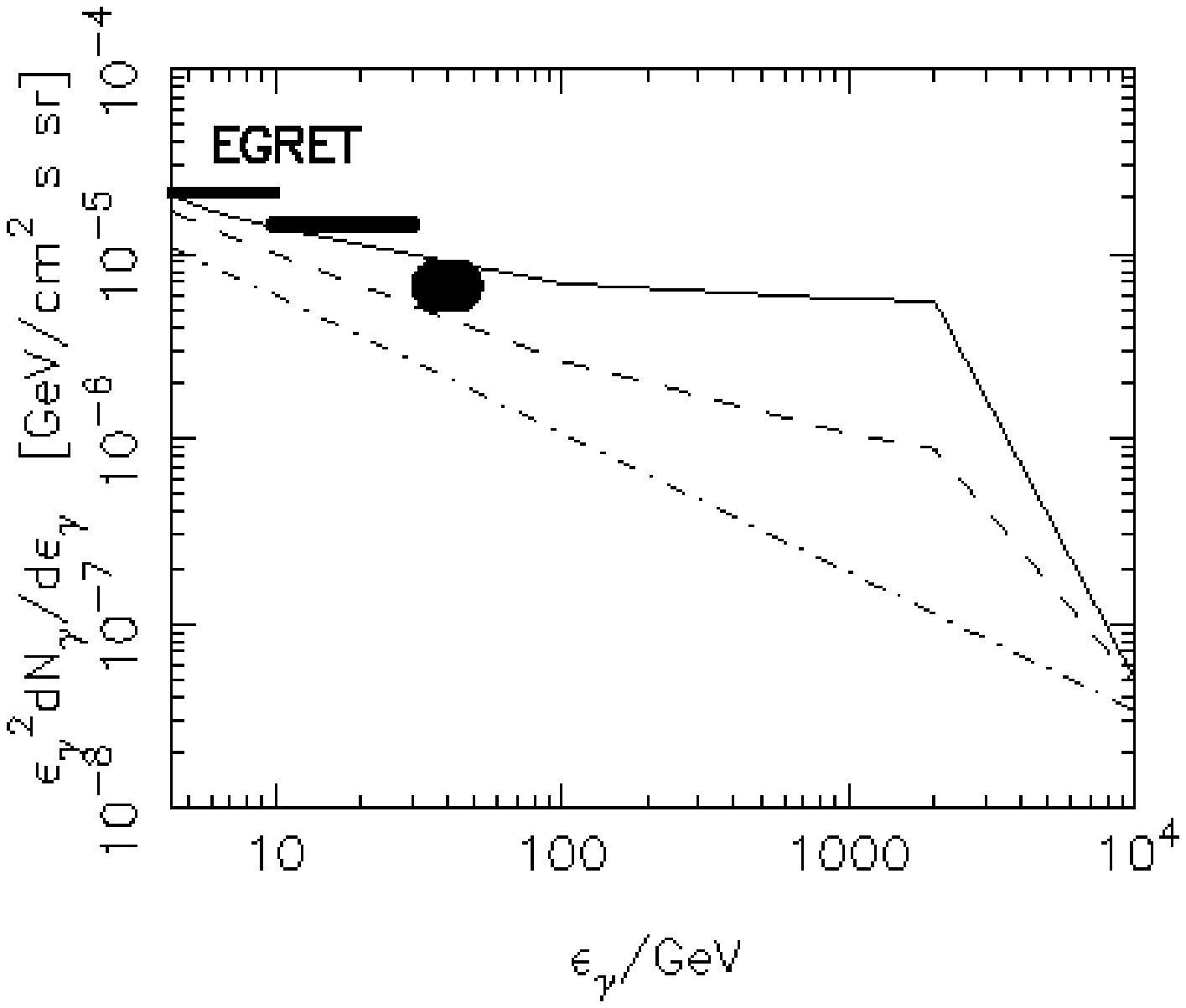}}
\caption{Integral diffuse $\gamma$-ray energy flux in the Galactic disk
(EGRET data from Hunter et al. 1997). The dash-dotted line is a model based
on the locally observed CR flux, whereas in the solid curve a purely
theoretical, and in the dashed curve a more realistic phenomenological SNR
source flux is added to this model flux. The fall-off beyond $\sim 2$~TeV is
due to the assumed escape of $\simge 20$~TeV CRs at late stages of source
evolution (from Berezhko \& V\"olk 2000).} 
\end{figure}

\begin{figure}[htbp]
\figbox*{}{}{\includegraphics*[width=8.2cm]{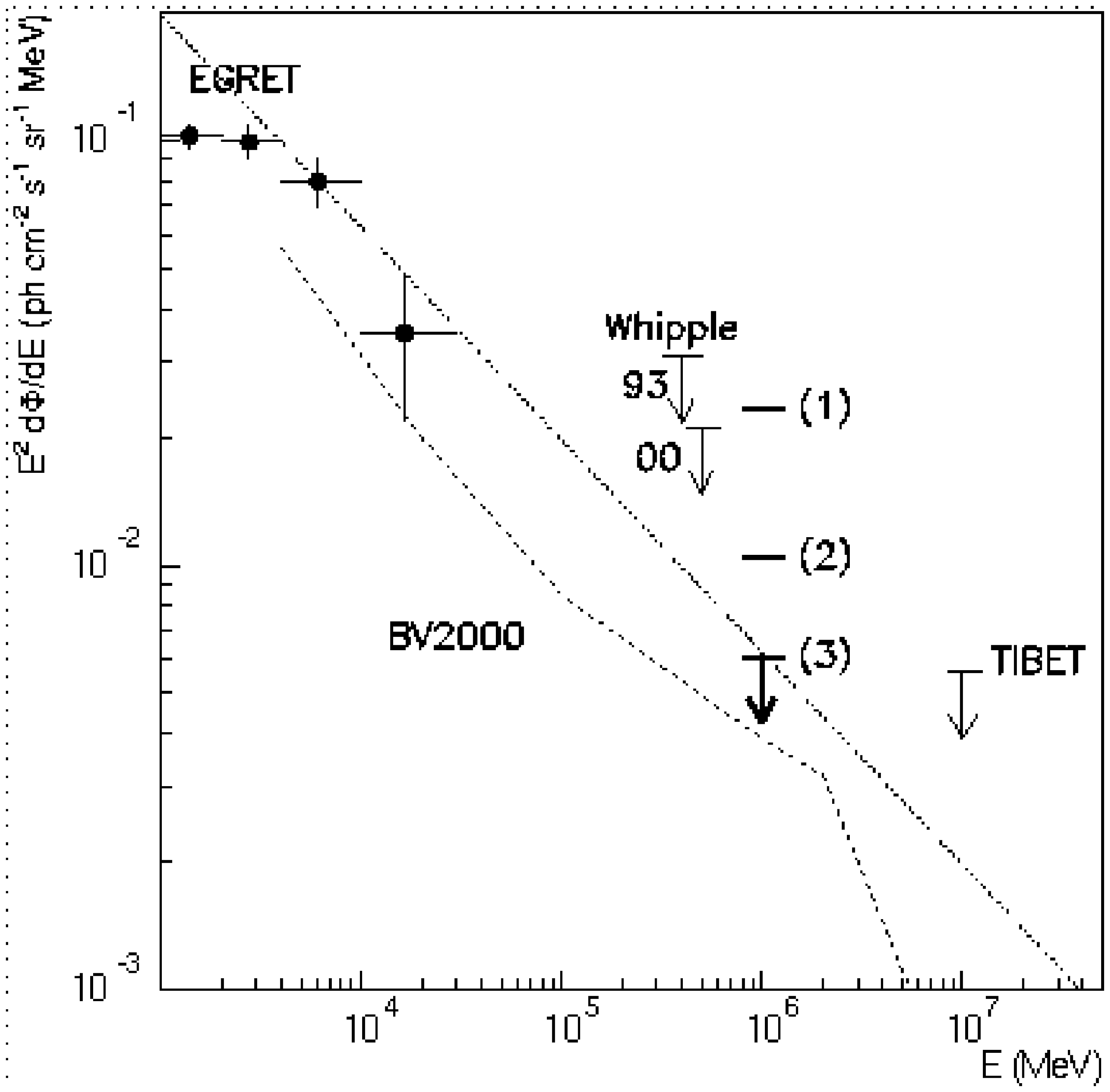}}
\caption{Integral diffuse $\gamma$-ray energy flux in the inner Galactic disk
together with TeV upper limits from several experiments compared to the scaled
prediction (cf. the dashed curve in Fig. 16), that includes unresolved SNR
sources. The lowest upper limit (3), from the HEGRA experiment, is quite close
to the model prediction (from Aharonian et al. 2001c).} 
\end{figure}

\subsection{The GeV excess}
\begin{sloppypar}
\noindent Extensive observations with the satellites SAS-II, Cos B, and CGRO in the range above
100~MeV have shown that the $\gamma$-ray flux from the Galactic disk is to lowest order consistent
with a uniform CR intensity there. This suggests effective spatial mixing, and is basically
consistent with the idea that CRs are diffusively confined in a large, quasi-spherical Galactic
Halo as put forward by the Moscow school (Ginz\-burg and Ptuskin, 1976). Such a measure of
uniformity for the CR density above 100 MeV suggests a rather uniform shape of the CR energy
spectrum as well. However, the EGRET instrument has found a hard $\gamma$-ray spectrum $\propto
E^{-2.4}$ at energies above 1~GeV which is in clear excess of the predicted spectrum $\propto
E^{-2.75}$, based on the locally observed CR nucleon spectrum.
\end{sloppypar}

Unexpected as it was, this discrepancy has raised a number of questions:

(i) Is the local CR nucleon spectrum not really representative
for the rest of the Galaxy on a large scale?

(ii) If this is indeed not the case, should we then for instance
think of an additional Inverse Compton emission from the neighborhood of
external sources of CR electrons?

(iii) Do we have to expect a substantial contribution $\propto
E^{-2.1}$ from the ensemble of unresolved CR sources?

\begin{figure}[htbp]
\figbox*{}{}{\includegraphics*[width=8.2cm]{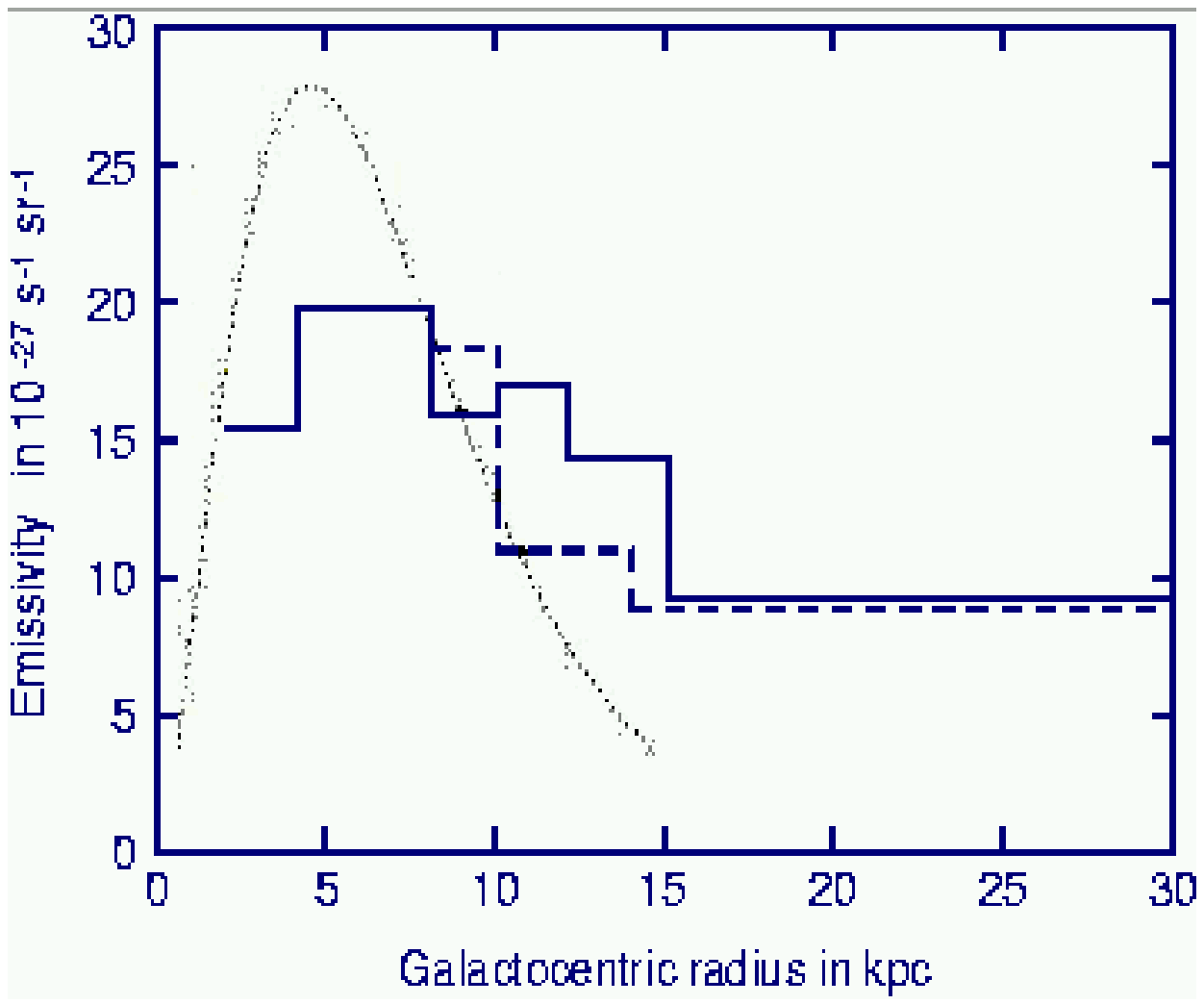}}
\caption{The $\gamma$-ray emissivity in the EGRET energy range as a function
of Galactocentric radius (solid and dashed histograms) in comparison to the
(dotted) SNR distribution as seen at radio frequencies (from Paul 2000).}
\end{figure}

Alternative (i) implies that along the $\gamma$-ray line of sight the
nucleon spectrum is different from the local one, possibly due to
propagation effects during escape. Given the fact that in a
Galactic Wind convection becomes stronger relative to diffusion with
increasing height above the disk midplane, most line of sights through
the disk should indeed include a hard-spectrum convective contribution.

As to the second question, it is true that high energy CR electrons can
propagate only over a rather limited range in the face of radiative losses.
Therefore, if the Solar system is not very close to a source, our local
electron spectrum may not be really representative at these higher
energies and possibility (ii) might contribute.

Perhaps the most interesting alternative is (iii). The estimated
hard-spectrum contribution of an unresolved SNR source distribution to the
"diffuse" $\gamma$-ray background in the disk is still small at 1 GeV but
finally dominant at TeV energies (Fig. 16). Thus the SNR sources should
"stick out" at high energies $E \sim $~TeV. Fig. 17 shows the latest
measurements of the TeV $\gamma$-ray background deep in the disk. Even
though not detected yet, its upper limit lies less than a factor 2 above
the expected flux. A clearly detected energy spectrum for $0.1 \simle E
\simle 1$~TeV should in the future allow a direct measurement of the
average CR source spectrum in the TeV range by the subtraction of a model
flux based on the locally observed CR nucleon spectrum from this diffuse
spectrum!

\begin{sloppypar}
Of course these alternatives are neither exclusive nor exhausting all possibilities. For example,
at energies $< 100$~GeV the unresolved source scenario could possibly still accomodate an
additional hard-spectrum component. However detection of angular fluctuations in the hard
$\gamma$-ray spectrum would at least separate the systematic effect (i) from the other two.
Ultimately, the detailed spectral investigation of a resolved background will be able to
distinguish cases (ii) and (iii).
\end{sloppypar}


\subsection{The average radial gamma-ray gradient in the Galaxy}

\noindent There is a second puzzling feature in the diffuse $\gamma$-ray
emission from the Galactic disk. At GeV energies the radial Ga\-lactocentric
$\gamma$~-ray emissivity gradient is too small to be explainable by uniform CR
diffusion from the SNR sources, which are assumed to lie in the disk (Fig. 18).
This might be interpreted as an argument against a CR origin from massive stars
in the disk since they are strongly concentrated in the $4$~to $6$~kpc molecular
ring around the center of the Galaxy. However, CR propagation away from the
Galaxy should actually be a strongly nonlinear process, determined by the
dynamical effects of the escaping CRs themselves. We must indeed expect that the
CRs from a large scale concentration of sources drive a faster Galactic Wind
from these sources by their enhanced pressure. This enhanced removal rate
counteracts the increase in the particle source strength in a self-regulating
manner (Fig. 19). Thus, spatial uniformity of CRs does not need to be exclusively
due to their diffusive nature, but as well due to the inability of the
Interstellar gas to weigh them down. Although considered here as an effect along
magnetic flux tubes directed away from the disk, such a behavior is basically
the result of a force balance requirement between gas, field and CRs, including
the gravity due to the stars.

\begin{figure}[t]
\figbox*{}{}{\includegraphics*[width=8.2cm]{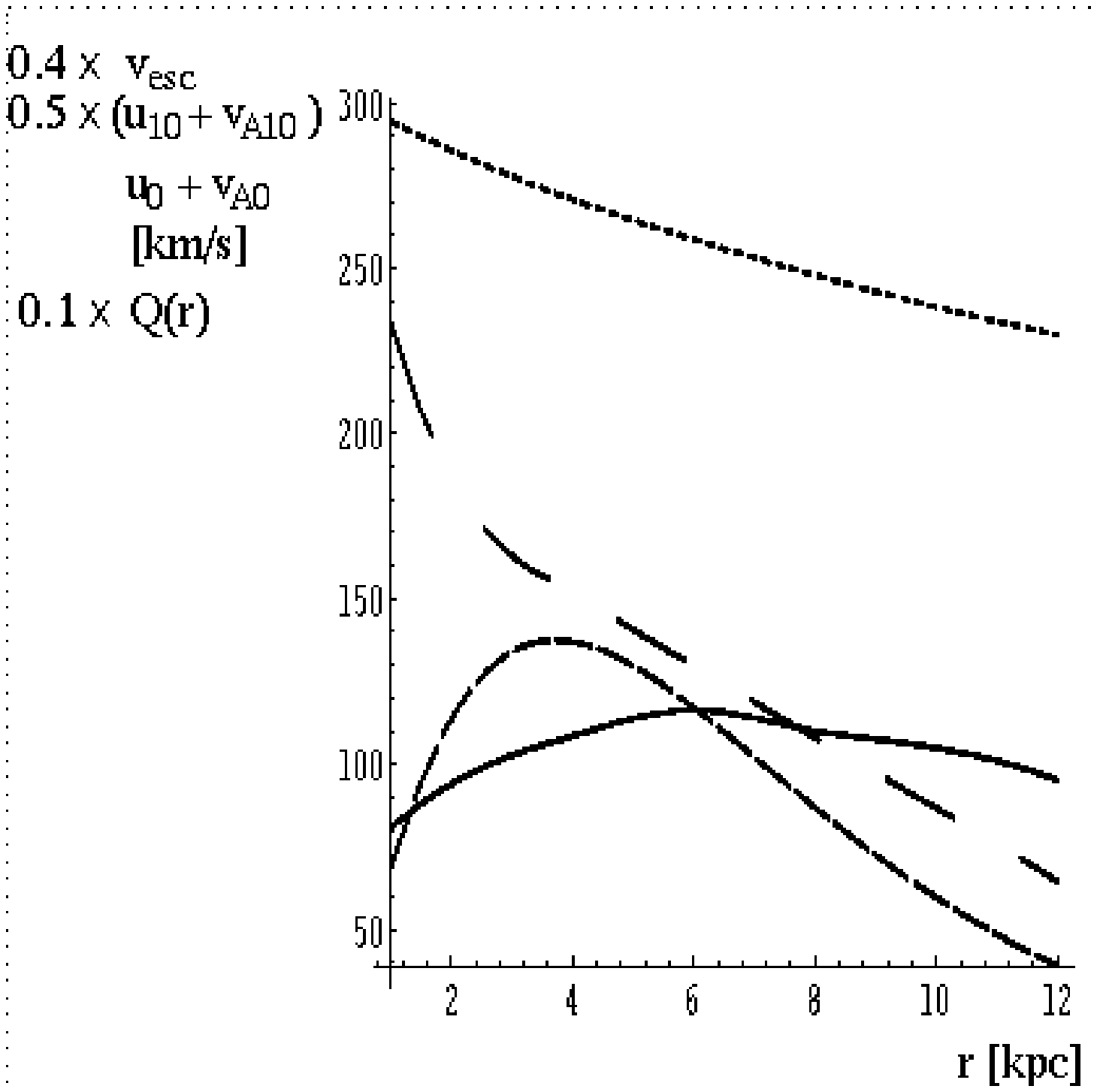}}
\caption{Escape speed $v_{esc}$ (dotted line) and CR drift velocity
$u_{10}+v_{A10}$ (long dashed line) at the large height of $10$~kpc above
the disk as a function of Galactocentric radius $r$; both fall off
radially due to the fall-off of the overall Galactic gravitational
potential distribution. The drift velocity $u_0+v_{A0}$ (solid line) at
the energy-averaged diffusion/convection boundary of the Galactic Wind at
$\simle 1$~kpc is more strongly influenced by the peak of the SNR
distribution, assumed to be proportional to the CR source distribution
(short dashed line). The source peak produces a maximum in $u_0+v_{A0}$ as
a function of radius and thus counteracts the CR density enhancement in a
steady state (from Breitschwerdt et al. 2001).}
\end{figure}

\begin{sloppypar}
An attractive feature of such a proposal is that it can be readily tested by the detection of the
diffuse $\gamma$-ray background in the TeV range: the ensemble of unresolved CR sources should
become visible directly at these $\gamma$-ray energies, reconstituting the radially peaked SNR
distribution that is so clearly seen at radio frequencies.
\end{sloppypar}

\section{Main forthcoming gamma-ray detectors}

\noindent The next years will see a number of new $\gamma$-ray detectors which are expected to
resolve the questions adressed in the two previous sections. The main projects are the Gamma-ray
LArge Satellite Telescope (GLAST) and several large ground-based telescopes. They are expected to
put high energy $\gamma$-astronomy on an entirely new level of flux sensitivity and wavelength
coverage with new classes of sources detected. Using the known gas density structures, especially
the distribution of Giant Molecular Clouds, the instruments should also be able to map the diffuse
CR density and energy distribution in the Galaxy in full detail.

\begin{figure}[htbp]
\figbox*{}{}{\includegraphics*[width=8.2cm,height=10cm]{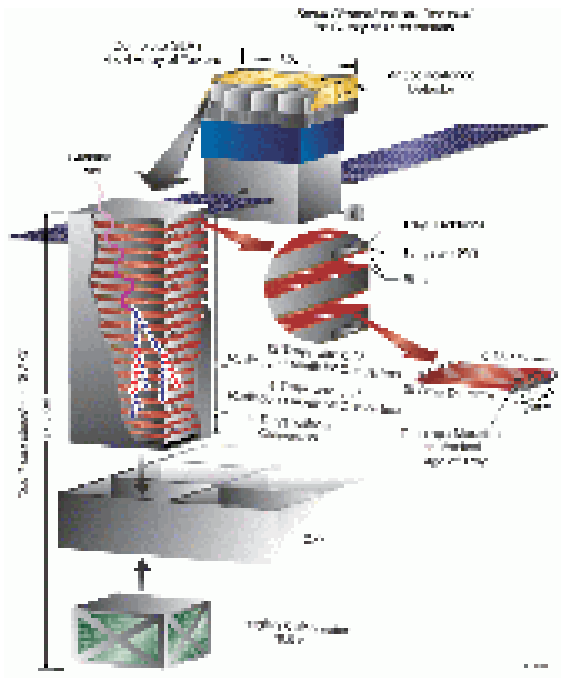}}
\caption{View of the GLAST instrument and its various detectors. The scheme
consists of a $4 \times 4$~array of so-called towers which contain converters for
$\gamma$-rays together with silicon trackers for the electron/positron
pairs, followed by an imaging calorimeter.} 
\end{figure}

\begin{figure*}[ht!]
\figbox*{}{}{\includegraphics*[width=17.cm,height=13.4cm]{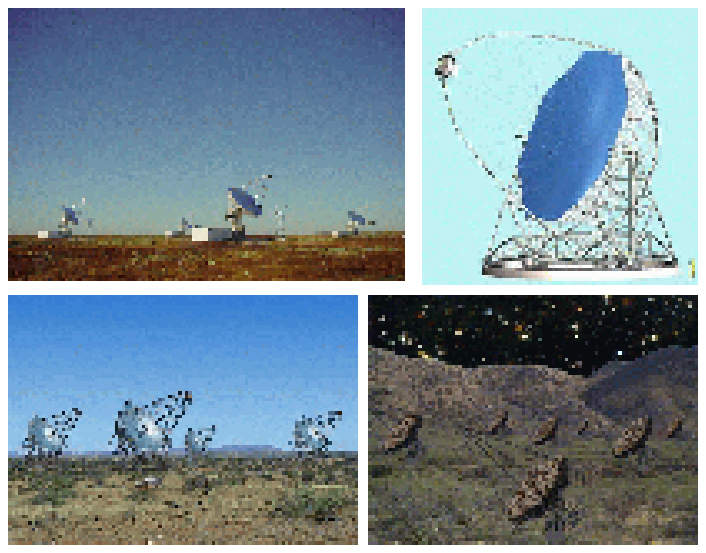}}
\caption{Photomontages/schematics of the main new Cherenkov telescopes. From the upper 
left to the lower right: the CANGAROO-III system of 4
telescopes (one telescope already in operation), the MAGIC telescope
(Major Atmospheric Gamma Imaging Cherenkov telescope), the
4-telescope Phase I of H.E.S.S. (High Energy Stereoscopic System), and
the 7-telescope array of VERITAS (Very Energetic Radiation Imaging
Telescope Array System).} 
\end{figure*}

NASA plans to launch GLAST in 2006. The instrument is based on silicon strip detectors and will
have a much improved angular resolution compared to EGRET - now matching that of the best ground
based instruments - and strongly increased sensitivity up to at least 100~GeV. At the higher
energies this sensitivity jump will in many regards remain rather limited by statistics, because
of the size restrictions on a detector operating on a satellite. In any case GLAST is expected to
survey the sky very effectively at lower energies due to its large field of view, complementary to
the higher threshold ground based arrays of telescopes whose major aim will be pointed
observations.

%

Four such large Cherenkov telescopes are under construction: CANGAROO
III in Australia, H.E.S.S. in Namibia, VERITAS in the U.S.A., and MAGIC on the
Canarian Island of La Palma (Fig. 21). The first three detectors are
stereoscopic systems like HEGRA, with telescope diameters of the order of
$10$~m, whereas MAGIC is a $17$~m diameter single telescope. The energy
thresholds are expected to lie between $30$ and $100$~GeV, and they will be
roughly an order of magnitude more sensitive than present instruments at $E
\sim 1$~TeV. The first telescope of CANGAROO III is already in operation,
H.E.S.S. and MAGIC are expected to see first light in early 2002, and VERITAS
plans for first light in 2003.

\subsection {The H.E.S.S. experiment in Namibia}

\noindent In conclusion let me say a few words about the H.E.S.S. array as a representative
example of the new ground based instruments. The name H.E.S.S. is an acronym for {\it High Energy
Stereoscopic System} and was chosen in honor of Victor Hess.

\begin{sloppypar}
The H.E.S.S. array\footnote{Detailed information on the experiment is to be found at the Web site:
http://www.mpi-hd.mpg.de/hfm/HESS/HESS.html.} is designed for coincident stereoscopic observations
with several telescopes. Phase I of the experiment consists of four $13$~m diameter telescopes, at
the corners of a square whose sides are $\approx 120$~m in length, roughly corresponding to the
radius of the Cherenkov light disk on the ground. The scientific collaboration involves about 60
individuals from European and Southern African institutions. The scientific and official Namibian
collaboration partner is the University of Namibia in Windhoek. The first telescope should become
operational by the beginning of 2002, the full Phase I system in 2003. H.E.S.S. is a natural
extension of the HEGRA stereoscopic system on La Palma and of the imaging technique in the CAT
telescope in the French Pyrenees.

\begin{figure}[t]
\figbox*{}{}{\includegraphics*[width=8.2cm,height=6cm]{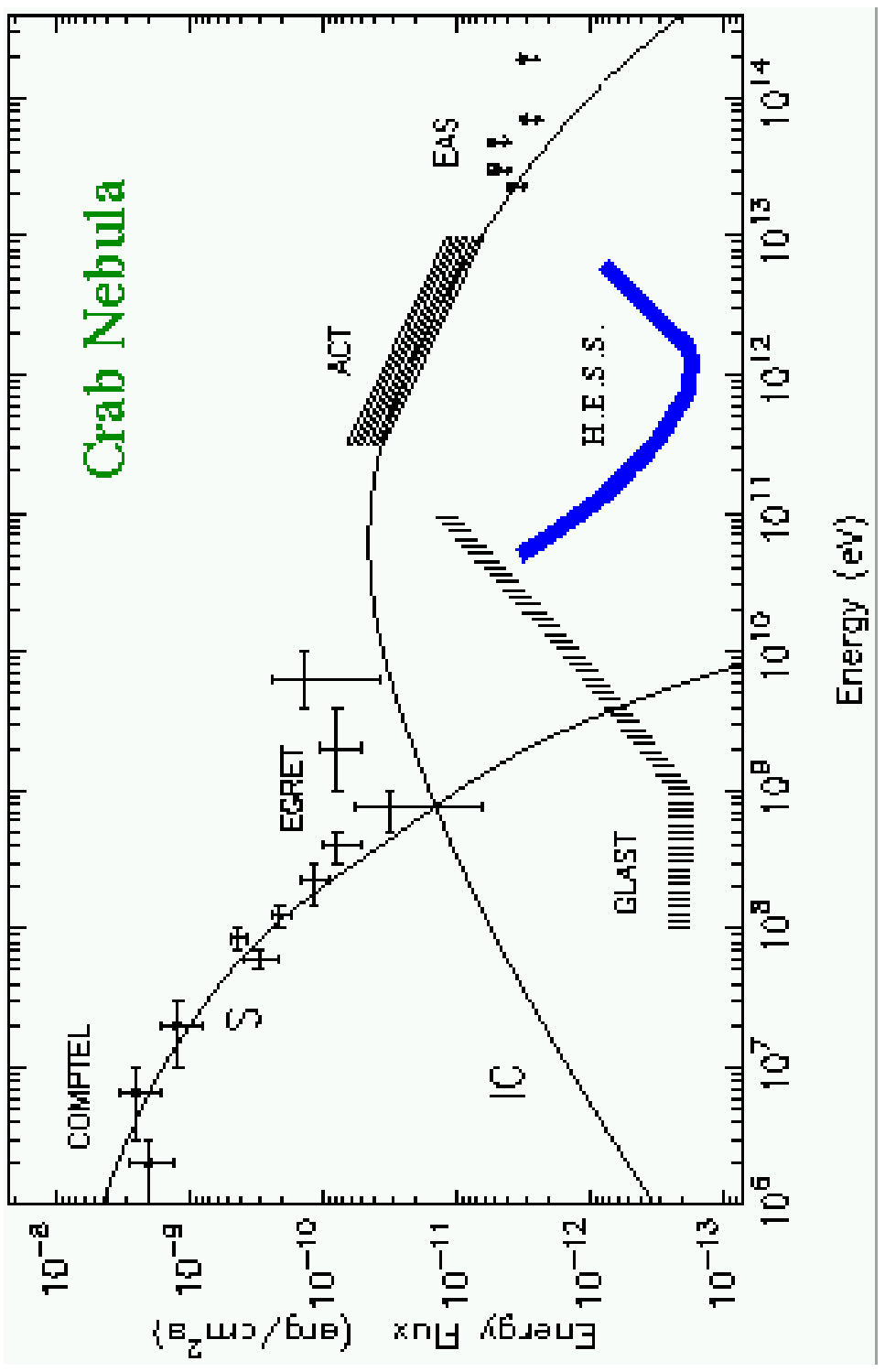}}
\caption{Performance of H.E.S.S. Using existing measurements of the Crab
Nebula together with model curves, for the spectral range $\simle 10^9$~eV
(synchrotron radiation) by the CGRO instruments COMPTEL and EGRET, and for
the range $\simge 10^{11}$~eV (Inverse Compton (IC) radiation) by imaging
atmospheric Cherenkov telescopes (ACT), as well as upper limits from
extensive air shower (EAS) arrays, the H.E.S.S. Phase I sensitivity for 50
hours of observation is plotted as a function of $\gamma$-ray energy. It is
also compared with the expectations for the GLAST instrument.}
\end{figure}

The H.E.S.S. telescopes will have a comparatively large field of view of $5$~degrees.  Their
energy threshold lies at $\approx 50$~ GeV for the detection of a source, and at $\approx 100$~GeV
for spectroscopic and spatially resolved observations. The angular and energy resolutions per
event are estimated as $0.1$~degrees, and 10 to 20 percent, respectively. The lowest energy flux
detectable in $50$~hours of observation time is about $10^{-12}$~erg/(cm$^2$s) above $100$~GeV,
and about $10^{-13}$~erg/(cm$^2$)s above $1$~TeV (Fig. 22). Therefore the hope is to find new
source populations whose TeV fluxes are about hundred times lower than that from the Crab Nebula.
\end{sloppypar}

\begin{figure}[htbp]
\figbox*{}{}{\includegraphics*[width=8.2cm]{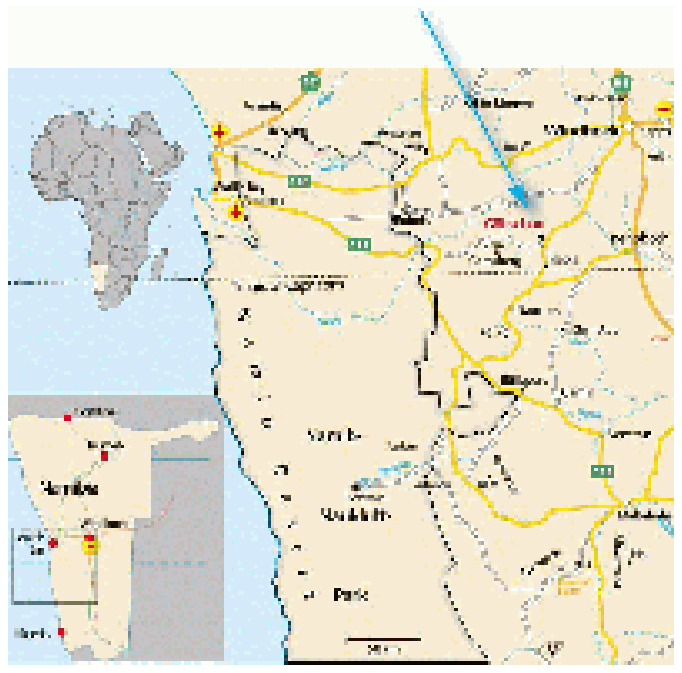}}
\caption{Section of a map of Namibia in southern Africa. The H.E.S.S. site
 (G\"ollschau) lies about $90$~km southwest of Windhoek.} 
\end{figure}

The site in Namibia is located in the Khomas Highland at $1800$~m above sea level. Geographically
this is almost precisely on the tropic of Capricorn, near the famous Gamsberg table mountain on a
$10$~km$^2$ piece of farm G\"ollschau (Fig. 23). It can be reached on a good dirt road from the
capital city of Windhoek in about $1.5$~hours. The Gamsberg area is one of the best optical sites
in the world and the mountain itself had been considered as a possible site for the European
Southern Observatory's Very Large Telescope project that is now on Cerro Paranal in Chile.


\begin{figure}[htbp]
\figbox*{}{}{\includegraphics*[width=8.cm]{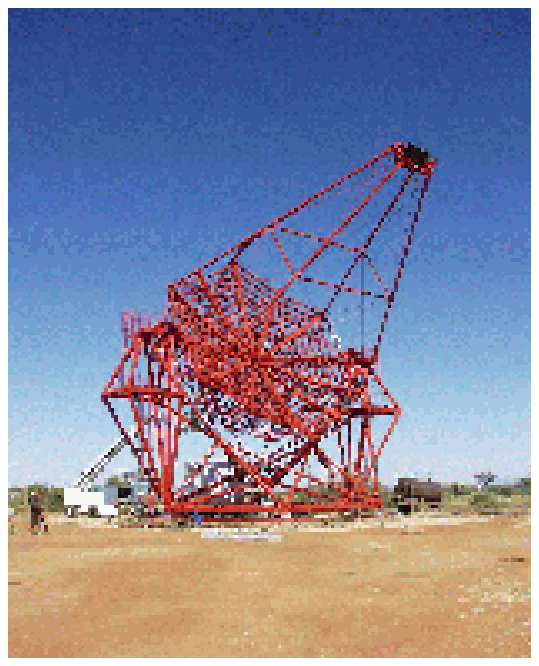}}
\caption{Steel frame of the first H.E.S.S. telescope. Rotation around 
the vertical axis is on a circular rail of $\sim 15$~m diameter.} 
\end{figure}

In the region cattle shares the scarce grass and water with occasional Kudus and Antilopes. The
steel frame of the first telescope is shown in Fig. 24. The 380 glass mirrors of $60$~cm diameter
each, and finally the focal plane detector (the "camera", Fig. 25) are due to be put on this frame
in fall of this year. Scientific operations of the telescope are expected to start in early 2002
if all goes well.
 
\begin{figure}[t]
\figbox*{}{}{\includegraphics*[width=8.2cm]{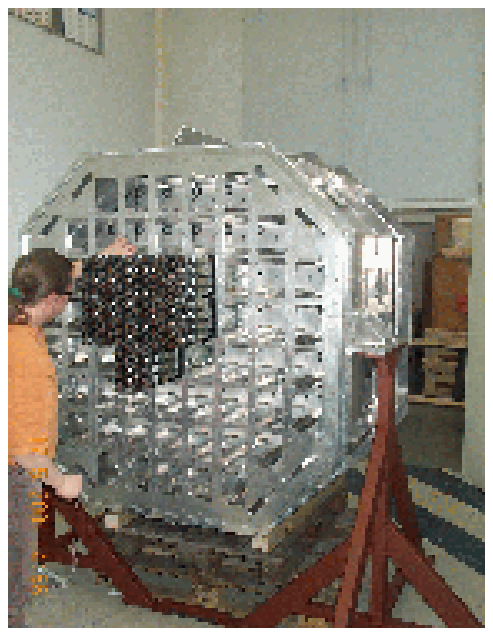}}
\caption{The first H.E.S.S. camera frame at the University VI -- VII in
Paris, in spring 2001 (with J.-P. Tavernet standing in front). At this
stage the camera was partly equipped with photomultipliers and the
corresponding electronics in the back. In full configuration the camera
has a total number of 980 phototubes (pixels) and a weight of about
$860$~kg.}
\end{figure}

\begin{acknowledgements}
  \begin{sloppypar}
I thank Alexander Konopelko for his comments on the manuscript.
  \end{sloppypar}
\end{acknowledgements}


\begin{thebibliography}{99}
\bibitem[1]{mor63}
Biographical information about VICTOR HESS, in: "Current Biography Yearbook
1963", (ed. Ch. Moritz), The H.W. Wilson Company, New York, p. 180--182,
1963.

\bibitem[1]{par66}
Parker, E.N., The Dynamical State of the Interstellar Gas and Field,
Astrophys. J., 145, 811--833, 1966.

\bibitem[2]{bre93} 
Breitschwerdt, D. McKenzie, J.F., V\"olk, H.J., Galactic winds. II. Role of
the Disk-Halo Interface in Cosmic Ray Driven Galactic Winds, A\&A 269,
54--66, 1993.

\bibitem[3]{ptu97} 
Ptuskin, V.S., V\"olk, H.J., Zirakashvili, V.N. et al., Transport of
Relativistic Particles in a Galactic Wind driven by Cosmic Rays, A\&A 321,
434--443, 1997.

\bibitem[4]{hes12} 
Hess, V.F., \"Uber Beobachtungen der durchdringenden Strahlung bei sieben
Freiballonfahrten, Physik. Zeitschr. 13, 1084, 1912.

\bibitem[5]{cro97}
Cronin, J.W., Gaisser, T.K., Swordy, S.P., Cosmic Rays at Energy Frontier,
Scientific American 276, 32--37, 1997.

\bibitem[6]{eng90}
Engelmann, J.J., Ferrando, P., Soutoul, A., et al., Charge composition and
energy spectra of cosmic-ray nuclei for elements from Be to Ni. Results from
HEAO-3-C2, A\&A 233, 96--111, 1990.

\bibitem[7]{swo90} 
Swordy, S.P., M\"uller, D., Meyer, P., et al., Relative Abundances, of
Secondary and Primary Cosmic Rays at High Energies, Astrophys. J. 349,
625--633, 1990.

\bibitem[6]{mal98} 
Malkov, M.A., V\"olk, H.J., Diffusive ion acceleration at
shocks: the problem of injection, Adv. Space Res. 21, No.4, 551--554, 1998.

\bibitem[6]{caw90} 
Cawley, M.F., Fegan, D.J., Harris, K., Hillas, A.M., et al., A high
resolution imaging detector for TeV gamma-ray astronomy, Experimental
Astronomy 1, 173--193, 1990.

\bibitem[6]{aha98}
Aharonian, F.A., Atoyan, A.M., Nonthermal Radiation of the Crab Nebula, in:
Proc. Int. Conf. on Neutron Stars and Pulsars (ed. Shibazaki, Kawai, Shibata,
Kifune), Tokyo, Frontiers Science Series No. 24, Universal Academic Press,
Inc., 439--448, 1998. (ISBN 4-946443-44-4)

\bibitem[6]{baa34}
Baade, W., Zwicky, F., Cosmic Rays from Supernovae, Proc Nat. Acad. Sci. USA
20, 259--263, 1934.

\bibitem[6]{tan98}
Tanimori, T., Hayami, Y, Kamei, S., et al., Discovery of TeV 
Gamma Rays from SN1006: Further Evidence for the SNR Origin of Cosmic Rays,
ApJ, 497, L25--L28, 1998.

\bibitem[6]{koy95}
Koyama, K., Petre, K., Gotthelf, E.V., et al., Evidence for shock
acceleration of high-energy electrons in the supernova remnant SN1006,
Nature, 378, 255--258, 1995.

\bibitem[6]{mur00}
Muraishi, H., Tanimori, T. Yanagita, S., et al., Evidence for TeV gamma-ray
emission from the shell type SNR RX J1713.7-3946, A\&A, 354, L57--L61, 2000.

\bibitem[6]{koy97}
Koyama, K., Kinugasa, K, Matsuzaki, K., et al., Discovery of non-thermal X-rays
from the northwest shell of the new SNR RX J1713.7-3946: the second SN 1006? Publ.
Astron. Soc. Japan, 49, L7--L11, 1997.

\bibitem[10]{dav94}
Drury, L.O'C., Aharonian, F.A. and V\"olk, H.J., The gamma-ray visibility
of supernova remnants. A test of cosmic ray origin, A\&A, 287, 959--971, 1994.

\bibitem[6]{nai94}
Naito, T., Takahara, F., High energy gamma-ray emission from supernova
remnants, J. Phys. G, 20, 477--486, 1994.

\bibitem[6]{aha01a}
Aharonian, F.A., Akhperjanian, A.G., Barrio, J.A. et al., Evidence for TeV
gamma-ray emission from Cassiopeia A, A\&A, 370, 112--120, 2001a.
  
\bibitem[1]{aha01b} 
Aharonian, F.A., Akhperjanian, A.G., Barrio, J.A. et al., A study of Tycho's
SNR at TeV energies with the HEGRA CT-system, A\&A, 373, 292--300, 2001b.

\bibitem[1]{gin76}
Ginzburg, V.L., Ptuskin, V.S., On the origin of cosmic rays: Some problems
in high-energy astrophysics, Rev. Mod. Phys. 48, 161--189, 1976.

\bibitem[1]{hun97}
Hunter, S.D., Kinzer, R.L., Strong, A.W., Diffuse Galactic Continuum
Radiation, AIP Conf. Proc. 410, Proc. Fourth Compton Symp., ed. Dermer, C.D.,
Kurfess, J.D. (New York: AIP), 192--207, 1997.

\bibitem[6]{bv00}
Berezhko, E.G., V\"olk, H.J., Galactic Gamma-Ray Background Radiation from
Supernova Remnants, ApJ 540, 923--929, 2000.

\bibitem[6]{aha01c}
Aharonian, F.A., Akhperjanian, A.G., Barrio, J.A. et al., A search for
gamma-ray emission from the Galactic plane in the longitude range between
$37^{\circ}$ and $43^{\circ}$, A\&A 375, 1008--1017, 2001c.

\bibitem[6]{pau01}
Paul, J., Gamma Rays from Star-Formation Regions, in: Int. Symp. on High Energy
Gamma-Ray Astronomy, Eds. F.A. Aharonian, H.J. V\"olk, Heidelberg, AIP Conf.
Proc., Melville, New York, Vol. 558, 183--190, 2001.

\bibitem[6]{bre01}
Breitschwerdt, D., Dogiel, V.A., V\"olk, H.J., The Gradient of diffuse
gamma-ray emission in the Galaxy. I. Insight from cosmic ray propagation
models, submitted to A\&A, 2001.

\end{thebibliography}
\end{document}